\documentclass[useAMS,usenatbib]{mn2e}
\usepackage{graphics}
\usepackage{epsf}
\usepackage{subfigure}
\def\ni{\noindent}

\begin{document}

\title[Cepheid PC \& AC relations II]{Period-colour and amplitude-colour relations in classical Cepheid variables II: the Galactic Cepheid models}
\author[Kanbur et al.]{Shashi M. Kanbur$^{1}$\thanks{E-mail: shashi@astro.umass.edu}, Chow-Choong Ngeow
$^{1}$ and J. Robert Buchler$^{2}$ 
\\
$^{1}$Department of Astronomy, University of Massachusetts      
\\
Amherst, MA 01003, USA
\\
$^{2}$Department of Physics, University of Florida,
\\
Gainesville, FL 32611, USA}

\date{Accepted 2004 month day. Received 2004 month day; in original form 2004 April 26}


\maketitle

\begin{abstract}
        In this paper, we construct full amplitude non-linear hydrodynamical models of fundamental mode Galactic Cepheids and analyze the resulting theoretical period-colour and amplitude-colour relations at maximum, mean and minimum light. These theoretical relations match the general form of the observed relations well. This agreement is, to some extent, independent of the mass-luminosity relations used, pulsation code, numerical techniques, details of the input physics and methods to convert theoretical quantities, such as bolometric luminosity and temperature, to observational quantities, such as V band magnitudes or $(V-I)$ colours. We show that the period-colour and amplitude-colour properties of fundamental mode Galactic Cepheids with periods such that $\log (P)>0.8$ can be explained by a simple application of the Stefan-Boltzmann law and the interaction of the photosphere with the hydrogen ionization front. We discuss the implications of our results for explaining the behavior of Galactic Cepheid period-colour, and period-luminosity relations at mean light.
\end{abstract}



\section{Introduction}

     \citet[][paper I, hereafter KN]{kan04} presented new observational characteristics of fundamental mode Cepheids obtained from an analysis of the Galactic Cepheid data and the OGLE LMC/SMC Cepheid data. These were the period-colour (PC) and the amplitude-colour (AC) relations at maximum, mean and minimum light. KN analyzed these PC and AC diagrams in the context of the work of \citet[][hereafter SKM]{sim93}, who used the Stefan-Boltzmann law and the fact that radius fluctuations are small in Cepheids, to derive the following equation, valid for optical pulsations of Cepheid variables:

     \begin{eqnarray}
       \log T_{max} - \log T_{min} = {1\over{10}}(V_{min} - V_{max}),
     \end{eqnarray}

\ni  where $T_{max/min}$ is the photospheric temperature at maximum or minimum light. Consequently, if for some reason the PC relation is flat at maximum light, and the colour used is a good predictor of temperature, then equation (1) predicts a relationship between $V$ band amplitude and $T_{min}$, and thus a correlation between the $V$ band amplitude and the observed colour (after extinction correction) at minimum $V$ band light. \citet{cod47} found that Galactic Cepheids exhibit a spectral type that is independent of period at maximum light. SKM analyzed existing Galactic Cepheid data to show that higher amplitude Galactic Cepheids are indeed driven to cooler temperatures, and thus redder $(B-V)$ colours, at minimum light in accordance with equation (1) and the observational findings of \citet{cod47}.

     The reason why Galactic Cepheids follow a spectral type that is independent of period at maximum light was explained in SKM: it is due to the interaction of the photosphere with the hydrogen ionization front (HIF). At maximum light, the HIF is so far out in the mass distribution that the photosphere, taken to be optical depth 2/3, occurs right at the base of the HIF. Together with the HIF is a sharp rise in opacity. At this point, the mean free path goes to zero. In the absence of any significant driving, even though the surrounding atmosphere has a non-zero inward velocity, this opacity wall prevents the photosphere from going deeper in the star and erases any ``memory'' of global stellar conditions. Thus for a large range of periods, the photosphere occurs at the base of the HIF at maximum light at a temperature which is independent of period. At maximum light, this leads to a flat relation between period \& temperature, period \& $(B-V)$ colour and period \& spectral type, as seen in SKM. At other phases, the HIF lies too far inside the mass distribution to interact with the photosphere. This, together with equation (1), implies a relation between amplitude and $(B-V)$ colour or temperature at minimum light. Equation (1) suggests that if $T_{min}$ obeyed a flat relation with period, then there should be a relation between amplitude and colour at maximum light. Hence PC relations are connected to AC relations, and changes in the PC relations should be reflected in corresponding changes in the AC relations. 

     Because of the extensive data now available, KN analyzed recent Galactic and Magellanic Cloud Cepheid data in terms of PC and AC diagrams at maximum, mean and minimum light. Using $(V-I)$ colours, they performed a more detailed analysis of the Galactic Cepheid data and found that the Galactic PC relation at maximum light displayed a statistically significant break at 10 days, but was consistent with a single line at mean and minimum light. In the LMC, the PC relation displayed this break at all three phases, though the statistical significance at maximum light was marginal. The SMC PC relation displayed similar properties to that in the Galaxy.

     Analysis of the Galactic Cepheid data in terms of AC diagrams confirmed the work of SKM, and extended it to the case of $(V-I)$ colour: there is a relation between amplitude and $(V-I)$ colour at minimum $V$ band light. For the LMC, short period Cepheids ($\log (P) < 1.0$) show no relation between $V$ band amplitude and $(V-I)$ colour at minimum light, but long period Cepheids ($\log (P) > 1.0$) are such that higher amplitude stars are driven to redder $(V-I)$ colours and hence cooler temperatures at minimum light. At maximum light, short period Cepheids are such that higher amplitude stars are driven to bluer $(V-I)$ colours and hence hotter temperatures but long period Cepheids do not show such a relation.

     An understanding of the PC and AC relations derived by KN are important, not only for stellar pulsation and evolution studies of Cepheids, but also because the Cepheid period-luminosity (PL) relation depends on the PC relation \citep[see, e.g., ][]{mad91}. The PL relation will reflect significant changes in the PC relation. Hence in studying PC and AC relations in different galaxies, we are studying the universality of the Cepheid PL relation. Recent work by \citet{tam02,tam03}, \citet{fou03}, \citet{kan04}, \citet{nge04}, \citet{san04} and \citet{sto04} have suggested that the PL relation in the Galaxy is significantly different from that in the LMC and, further, that the PL relation in the LMC is non-linear. In contrast, current observations indicate that the Galactic PL relation is linear \citep{tam03,kan04,nge04}.

     The purpose of this and subsequent papers is to confront these new observed characteristics of Cepheids with the latest stellar pulsation models and interpret the results in terms of the theory presented in SKM, which has been summarized above. This approach will ultimately yield a qualitatively deeper understanding some of the reasons behind the variation of the Cepheid PL relation from galaxy to galaxy. Our ultimate goal is to use our theoretical models to estimate the effect of this variation quantitatively. For this paper we concentrate on Galactic Cepheid models which we compare with the same Galactic Cepheid data used in KN. The LMC/SMC Cepheid models will be presented in the forthcoming papers in this series.

     Our models and methods improve upon those in SKM in the following respects: First of all, they contain a formulation to model time dependent turbulent convection \citep{yec99,kol02}, in contrast to the purely radiative models used in SKM. In addition, we construct more models so that we can examine in greater detail the PC and AC characteristics of Cepheids. Secondly, we investigate the pulsation properties as a function of other phase points. When investigating PC and AC relations at mean light, KN defined mean $(V-I)$ colour as $V_{mean}-I_{phmean}$ where $V_{mean}$ is the $V$ band magnitude closest to the value of $A_0$, the mean magnitude obtained from a Fourier decomposition, and $I_{phmean}$ is the $I$ band magnitude at this same phase. The reason for doing this is because the colour of the star at a certain phase is precisely the $V$ band magnitude minus the $I$ band magnitude at that same phase. In contrast, a definition of mean colour such as $<V> - <I>$ folds in the phase difference (albeit small) that exists between the $V$ and $I$ band light curves. We adopt the definitions given in KN for colour at maximum, mean and minimum light. However, there are two phases when the $V$ band magnitude is closest to the value of $A_0$: once on the ascending branch and once on the descending branch of the light curve. The colour (or the temperature), in both models and theory, need not necessary be the same at these two phases. In KN's analysis of the observed data, the descending branch mean was adopted as the ``mean''. We pay specific attention to this detail in our results and discussion section. Finally, we look at $(V-I)$ colours whereas SKM studied predominantly $(B-V)$ colours. The $(V-I)$ colour is a good indicator of temperature \citep{gon96,tam03} and is a crucial colour for the existing calibration of the extra-galactic distance scale (e.g., \citealt{fre01}).
     

\begin{table}
  \centering
  \caption{Input parameters for Galactic Cepheid models with periods obtained from a linear analysis. The periods, $P_0$ and $P_1$, are referred to the fundamental and first overtone periods, respectively. Similarly for the growth rate, $\eta$. Both of the mass and luminosity are in Solar unit, the temperature is in $K$ and the period is in days.}
  \label{tab1}
  \begin{tabular}{ccccccc} \hline
    M & $\log(L)$ & $T_{e} $ & $P_0$ & $\eta_0$ & $P_1$ & $\eta_1$ \\
    \hline \hline
    \multicolumn{7}{c}{ML Relation from \citet{bon00}}
    \\ \hline
    11.5 & 4.279 & 4830 & 43.3924 & 0.042 & 27.58 & -0.164 \\
    11.1 & 4.228 & 4900 & 37.8090 & 0.039 & 24.40 & -0.157 \\
    10.8 & 4.188 & 4950 & 34.4156 & 0.037 & 22.25 & -0.133 \\
    10.5 & 4.147 & 5000 & 30.8189 & 0.035 & 20.26 & -0.119 \\
    9.55 & 4.009 & 5075 & 23.4835 & 0.025 & 15.69 & -0.094 \\
    9.45 & 3.994 & 5265 & 20.0414 & 0.028 & 13.55 & -0.051 \\
    8.60 & 3.856 & 5300 & 15.8313 & 0.022 & 10.78 & -0.040 \\
    7.70 & 3.696 & 5332 & 12.1076 & 0.015 & 8.291 & -0.036 \\
    7.30 & 3.618 & 5440 & 10.0124 & 0.014 & 6.886 & -0.017 \\
    7.30 & 3.618 & 5490 & 9.69213 & 0.014 & 6.670 & -0.007 \\
    7.00 & 3.557 & 5490 & 8.83394 & 0.013 & 6.087 & -0.010 \\ 
    6.80 & 3.515 & 5485 & 8.31328 & 0.011 & 5.733 & -0.014 \\
    6.45 & 3.438 & 5545 & 7.12090 & 0.011 & 4.921 & -0.007 \\
    6.10 & 3.357 & 5580 & 6.16131 & 0.009 & 4.267 & -0.006 \\
    6.00 & 3.333 & 5590 & 5.59661 & 0.006 & 3.884 & -0.007 \\
    \hline 
    \multicolumn{7}{c}{ML Relation from \citet{chi89}}
    \\ \hline
    10.0 & 4.534 & 5150 & 66.9094 & 0.144 & 38.47 & -0.413 \\
    8.50 & 4.279 & 5100 & 44.8456 & 0.129 & 27.38 & -0.238 \\
    8.00 & 4.184 & 5090 & 38.5673 & 0.116 & 23.98 & -0.196 \\
    7.40 & 4.061 & 5050 & 32.4902 & 0.096 & 20.55 & -0.171 \\
    7.20 & 4.019 & 5250 & 26.1109 & 0.081 & 16.99 & -0.115 \\
    7.00 & 3.975 & 5400 & 21.8715 & 0.060 & 14.45 & -0.077 \\
    6.30 & 3.810 & 5390 & 16.9148 & 0.055 & 11.37 & -0.048 \\
    5.50 & 3.597 & 5350 & 12.4229 & 0.038 & 8.445 & -0.051 \\
    5.10 & 3.478 & 5396 & 10.0049 & 0.030 & 6.851 & -0.041 \\
    5.00 & 3.447 & 5420 & 9.37928 & 0.028 & 6.436 & -0.036 \\
    4.90 & 3.416 & 5440 & 8.80955 & 0.026 & 6.056 & -0.033 \\
    4.80 & 3.383 & 5470 & 8.21323 & 0.025 & 5.657 & -0.027 \\
    4.50 & 3.282 & 5460 & 7.06755 & 0.017 & 4.878 & -0.037 \\
    4.40 & 3.247 & 5490 & 6.56235 & 0.016 & 4.536 & -0.032 \\
    4.20 & 3.174 & 5560 & 5.60322 & 0.015 & 3.887 & -0.022 \\
    \hline 
    \multicolumn{7}{c}{Input parameters from SKM}
    \\ \hline
    4.57 & 3.306 & 5707 & 10.5472 & 0.035 & 7.234 & -0.000 \\
    5.44 & 3.578 & 5550 & 6.25949 & 0.022 & 4.340 &  0.020 \\
    \hline 
  \end{tabular}
\end{table}



\begin{table}
  \centering
  \caption{Temperatures at maximum and minimum light from full-amplitude non-linear model calculations. The periods, luminosity and temperature are in days, $L_{\odot}$ and $K$, respectively. \label{tab2}}
  \begin{tabular}{ccccc} \hline
    $P$ & $L_{max}$ & $T_{max}$ & $L_{min}$ & $T_{min}$ \\ 
    \hline \hline
    \multicolumn{5}{c}{ML Relation from \citet{bon00}}
    \\ \hline
    43.3924    & 23323.69 & 5301.44 & 15265.45 & 4653.25 \\  
    37.8090    & 20547.96 & 5355.92 & 13844.78 & 4759.77 \\
    34.4156    & 18675.64 & 5399.02 & 12811.00 & 4681.60 \\
    30.8189    & 17046.03 & 5446.49 & 11592.02 & 4719.65 \\
    23.4835    & 12259.96 & 5456.64 & 8535.450 & 4820.82 \\
    20.0414    & 12503.63 & 5720.76 & 7897.603 & 4948.82 \\ 
    15.8313    & 8828.696 & 5664.44 & 6018.911 & 5025.59 \\
    12.1076    & 5616.756 & 5525.95 & 4402.925 & 5135.11 \\
    10.0124    & 4470.551 & 5531.94 & 3703.182 & 5274.68 \\
    9.69213    & 4484.612 & 5585.30 & 3647.089 & 5303.57 \\
    8.83394    & 3872.188 & 5619.60 & 3172.488 & 5312.02 \\
    8.31328    & 3501.343 & 5612.30 & 2906.126 & 5322.39 \\
    7.12090    & 2954.314 & 5673.40 & 2436.595 & 5396.37 \\
    6.16131    & 2434.291 & 5700.83 & 2059.419 & 5459.73 \\
    5.59661    & 2260.885 & 5691.47 & 2023.329 & 5513.33 \\
    \hline 
    \multicolumn{5}{c}{ML Relation from \citet{chi89}}
    \\ \hline
    66.9094    & 39933.45 & 5468.07 & 25451.85 & 4807.32 \\
    44.8456    & 24179.71 & 5738.10 & 12478.80 & 4707.13 \\               
    38.5673    & 19927.26 & 5762.38 & 9964.144 & 4696.57 \\
    32.4902    & 15175.07 & 5724.16 & 7818.562 & 4720.46 \\
    26.1109    & 13968.71 & 5886.81 & 7294.068 & 4949.57 \\ 
    21.8715    & 13020.35 & 6056.17 & 7076.027 & 5176.93 \\
    16.9148    & 8608.982 & 5999.40 & 4782.056 & 4961.41 \\
    12.4229    & 5070.216 & 5846.81 & 3081.266 & 4998.02 \\
    10.0049    & 3724.848 & 5785.06 & 2465.684 & 5074.73 \\
    9.37928    & 3427.493 & 5770.41 & 2320.390 & 5115.94 \\ 
    8.80955    & 3130.614 & 5748.80 & 2187.659 & 5156.00 \\
    8.21323    & 2843.303 & 5729.82 & 2062.474 & 5217.61 \\ 
    7.06755    & 2080.732 & 5572.38 & 1722.053 & 5295.72 \\
    6.56235    & 1885.999 & 5565.48 & 1587.056 & 5332.63 \\
    5.60322    & 1580.301 & 5697.21 & 1315.624 & 5383.56 \\ 
    \hline 
    \multicolumn{5}{c}{Input parameters from SKM}
    \\ \hline
    10.5472    & 2276.417 & 5894.64 & 1660.255 & 5430.83 \\
    6.25949    & 5084.652 & 6068.32 & 2974.031 & 5170.89 \\
    \hline 
  \end{tabular}
\end{table}


\section{Models and Code description}

     The numerical techniques and physics included in these models are detailed in \citet{yec99} and \citet{kol02}. The Florida code has been used successfully to model double mode Cepheid pulsations \citep{kol98}, Galactic first overtone Cepheids \citep{feu00}, Cepheid mass-luminosity (ML) relations \citep{bea01} and the study of Magellanic Clouds Cepheids \citep{buc04}. The code takes the mass, luminosity, effective temperature, hydrogen and metallicity abundance (by mass), $M, L, T_e, X, Z$, as input parameters. In this study, we used $(X,Z)=(0.70,0.02)$ as a representation of Galactic hydrogen and metallicity abundance. For the remaining parameters, mass and luminosity are connected by an adopted ML relation, hence it is only necessary to choose the mass and the effective temperature for computing a Cepheid model. In this study, we used two ML relations, which are calculated from evolutionary models appropriate for intermediate-mass stars:

     \begin{enumerate}
     \item  The ML relation given by \citet{chi89}, which was also used by \citet{sim95} in their theoretical study of long period Cepheids:
       \begin{eqnarray}
         \log L = 3.61 \log M + 0.924.
       \end{eqnarray}
       
     \item  The ML relations given by \citet{bon00}:
       \begin{eqnarray}
         \log L & = & 0.90 + 3.35 \log M + 1.36\log Y - 0.34 \log Z, \\
         \nonumber & = & 3.35 \log M + 0.726.       
       \end{eqnarray}
     \end{enumerate}

\ni The units for both $M$ and $L$ are in Solar unit. Note that equation (2) gives a much larger $L/M$ ratio than equation (3). The range of $L/M$ ratio covered by equations (2) and (3) is reasonably broad and encompasses a fair selection of ML relations in the literature. Our aim for this paper is not to test one ML relation against another, but rather, to show that the physical effect we are interested in is, to a large extent, independent of the ML relation chosen.

     The effective temperature, $T_e$, was chosen to ensure a good fundamental growth rate and a stable first overtone mode. After a linear non-adiabatic analysis \citep{yec99} which yields the normal mode spectrum, an initial perturbation, which is scaled to match the fundamental mode linear velocity eigenvector, is applied and the model followed until a stable limit cycle is reached. In some cases, this means following the pulsations for about 1000 cycles. This results in a full amplitude Cepheid oscillating in the fundamental mode. Following SKM, we can compute the temperature at an optical depth $\tau=2/3$ and also the temperature and opacity profiles as a function of depth at various phases of the pulsation. 

     In addition to the two ML relations used, we have also computed two models using the parameters ($M,L,T_e$) from table 1 of SKM, in order to compare the results based on purely radiative and the turbulent convective models, and further broaden the range of ML relation considered. The input parameters for these models and for the two ML relations used are presented in Table \ref{tab1}. 

     We compare theoretical quantities to observable quantities using a number of prescriptions:
     \begin{enumerate}
     \item  We convert observed, extinction corrected $(V-I)$ to $(B-V)$ colours using the formula given in \citet{tam03}, and then convert the $(B-V)$ index to temperatures using the formula given in SKM.

     \item  We use the BaSeL atmosphere database\footnote{\texttt{http://www.astro.mat.uc.pt/BaSeL/}} \citep{lej02,wes02} to construct a fit giving temperature as a function of $(V-I)$ colour, with and without a $\log (g)$ term. Where a $\log (g)$ term is used, we obtain this, at the appropriate phase, from the models. This database is also used in, for example, \citet{bea01} and \citet{cor03} for the study of Magellanic Cloud Cepheids. 

     \item  These techniques were also used to obtain the bolometric corrections required to convert theoretical luminosity curves into theoretical $V$ band light curves. \citet{sim95} have shown that these corrections are small.
 
     \end{enumerate}

     The qualitative thrust of our results does not change according to the details of the method used to convert colours to temperatures and luminosity to $V$ band light. In what follows, we report on results using method (ii) above with a $\log (g)$ term. One caveat here is that we use hydrostatic equilibrium atmospheres. The effective temperature and effective gravity for these atmospheres are obtained from the relevant quantities at the photosphere in the models. For the effective gravity we use

     \begin{eqnarray}
     \nonumber g & = &\frac{GM}{R^2} - \frac{du}{dt}, 
     \end{eqnarray}

\ni where $du/dt$ is again given by the instantaneous structure of the models. Such an approach has been widely used in the literature in recent years for both Cepheids and RR Lyraes (see, for examples, SKM, \citealt{sim95}, \citealt{bon99}, \citealt{san99}, \citealt{bea01}, \citealt{cap02}, \citealt{kov03}, \citealt{rou04}). Some justification based on detailed modeling may be found in \citet{kol00} and \citet{kel70}. We note that at most phases the $du/dt$ term is small but can become large at certain phases during the ascending and descending branches. For the models considered here, this occurs for three models with $\log (P)$ between 1.5 and 1.65 at phases close to maximum light.


     \begin{figure*}
       \vspace{0cm}
       \hbox{\hspace{0.2cm}\epsfxsize=8.5cm \epsfbox{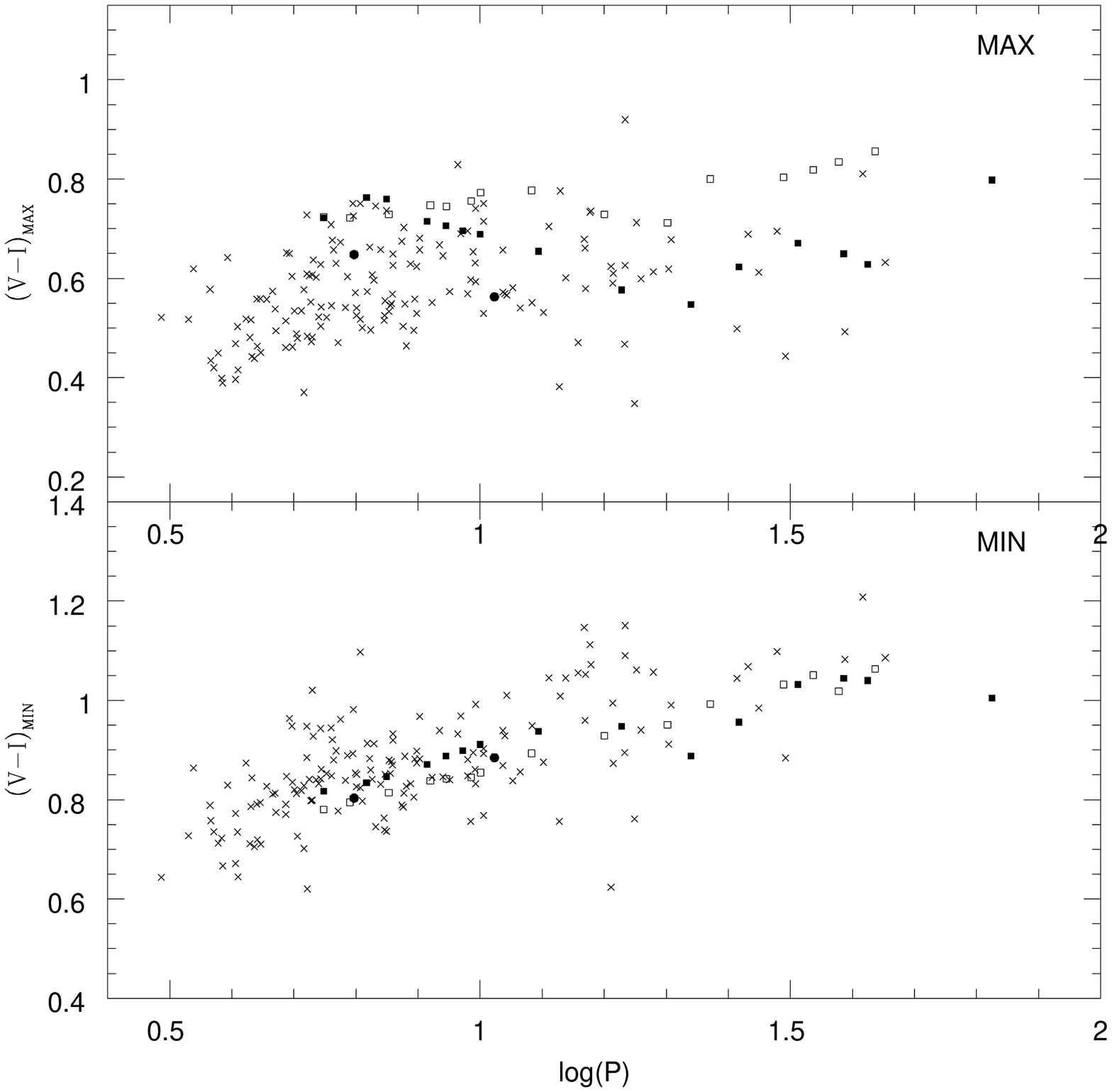}
         \epsfxsize=8.5cm \epsfbox{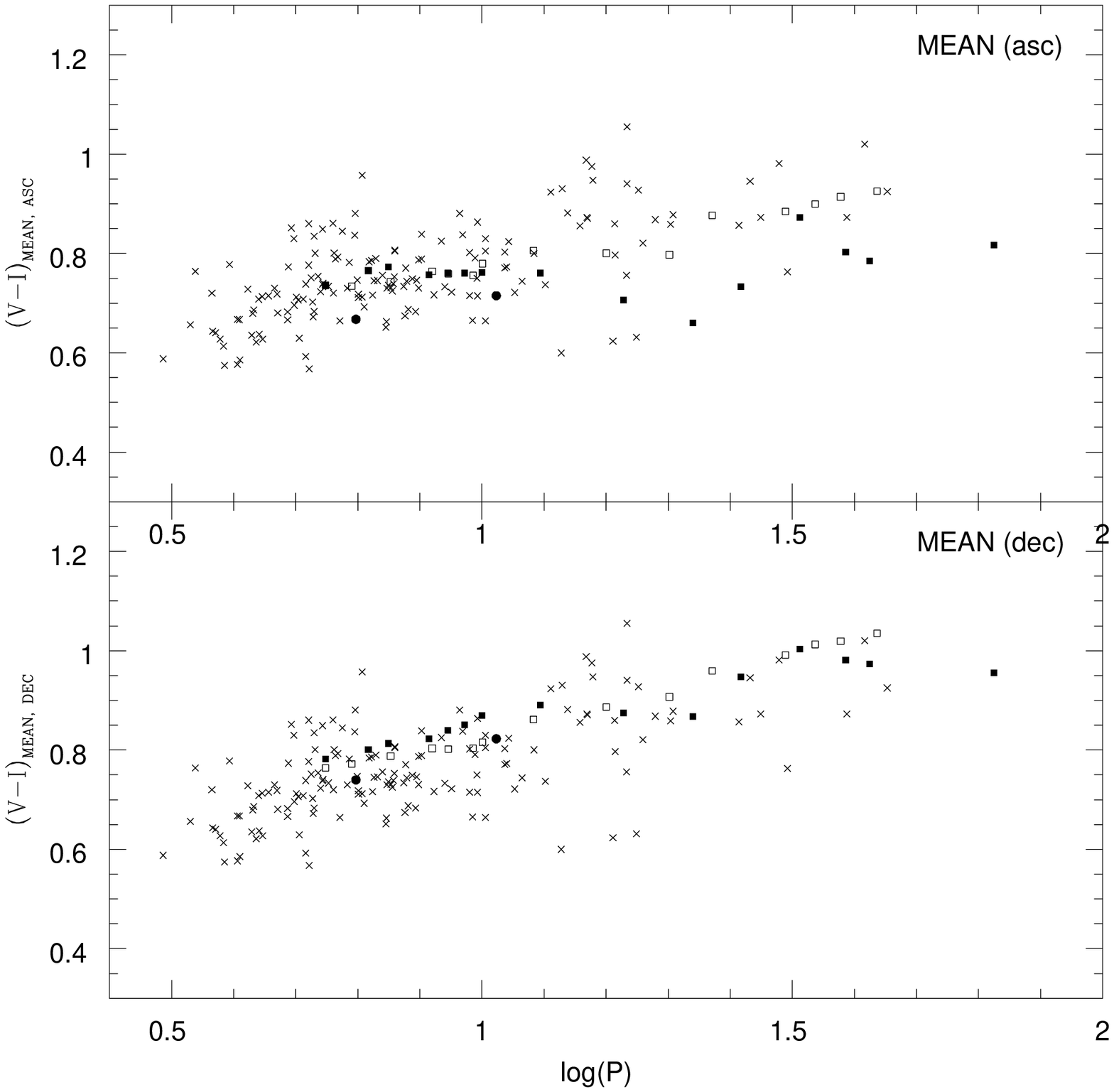}}
       \vspace{0cm}
       \caption{The period-colour (PC) relations for the Galactic Cepheids. The crosses are the observed data points, which are taken from \citet{kan04}. The open and solid squares are the Galactic models with \citet{bon00} and \citet{chi89} ML relations, respectively, whilst the solid circles are results for models computed with the ML relation used in SKM. The temperatures from the models are converted to the $(V-I)$ colour using the BaSeL database with the $\log(g)$ term. {\it Left} (a): PC relations at maximum and minimum light; {\it Right} (b): PC relation at mean light, for both of the acceding and descending branches. \label{fig1}}
     \end{figure*}

     
     \begin{figure*}
       \vspace{0cm}
       \hbox{\hspace{0.2cm}\epsfxsize=8.5cm \epsfbox{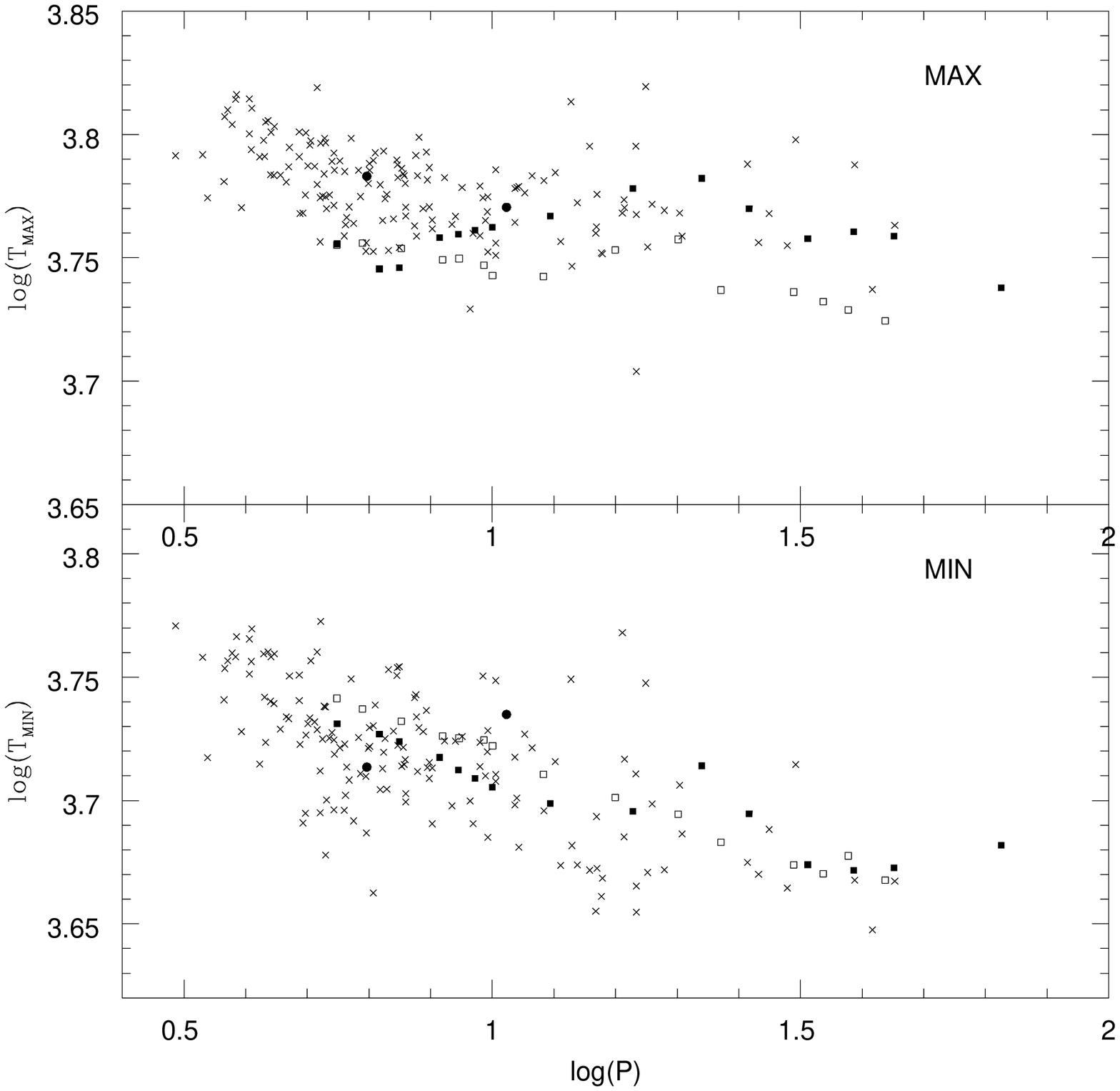}
         \epsfxsize=8.5cm \epsfbox{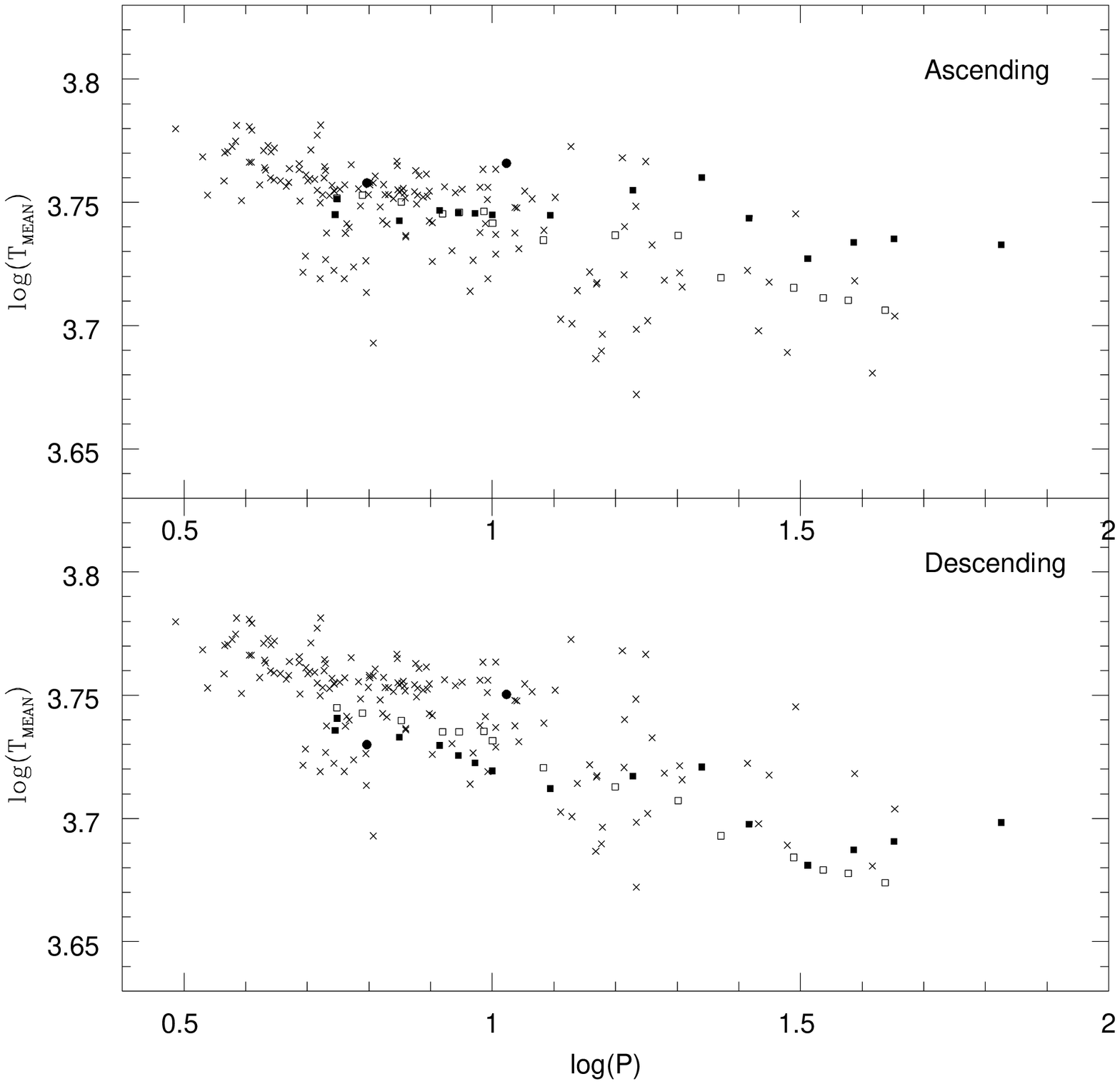}}
       \vspace{0cm}
       \caption{The plots of $\log(T)$ vs. $\log(P)$ for the Galactic data and models. The symbols are same as in Figure \ref{fig1}. The conversion from observed $(V-I)$ colour to temperature is done with the BaSeL database. {\it Left} (a): The $\log(P)$-$\log(T)$ plots at maximum and minimum light; {\it Right} (b): The $\log(P)$-$\log(T)$ plots at mean light, for both of the ascending and descending branches. \label{fig2}}
     \end{figure*}

 
\begin{figure*}
  \vspace{0cm}
  \hbox{\hspace{0.2cm}\epsfxsize=8.5cm \epsfbox{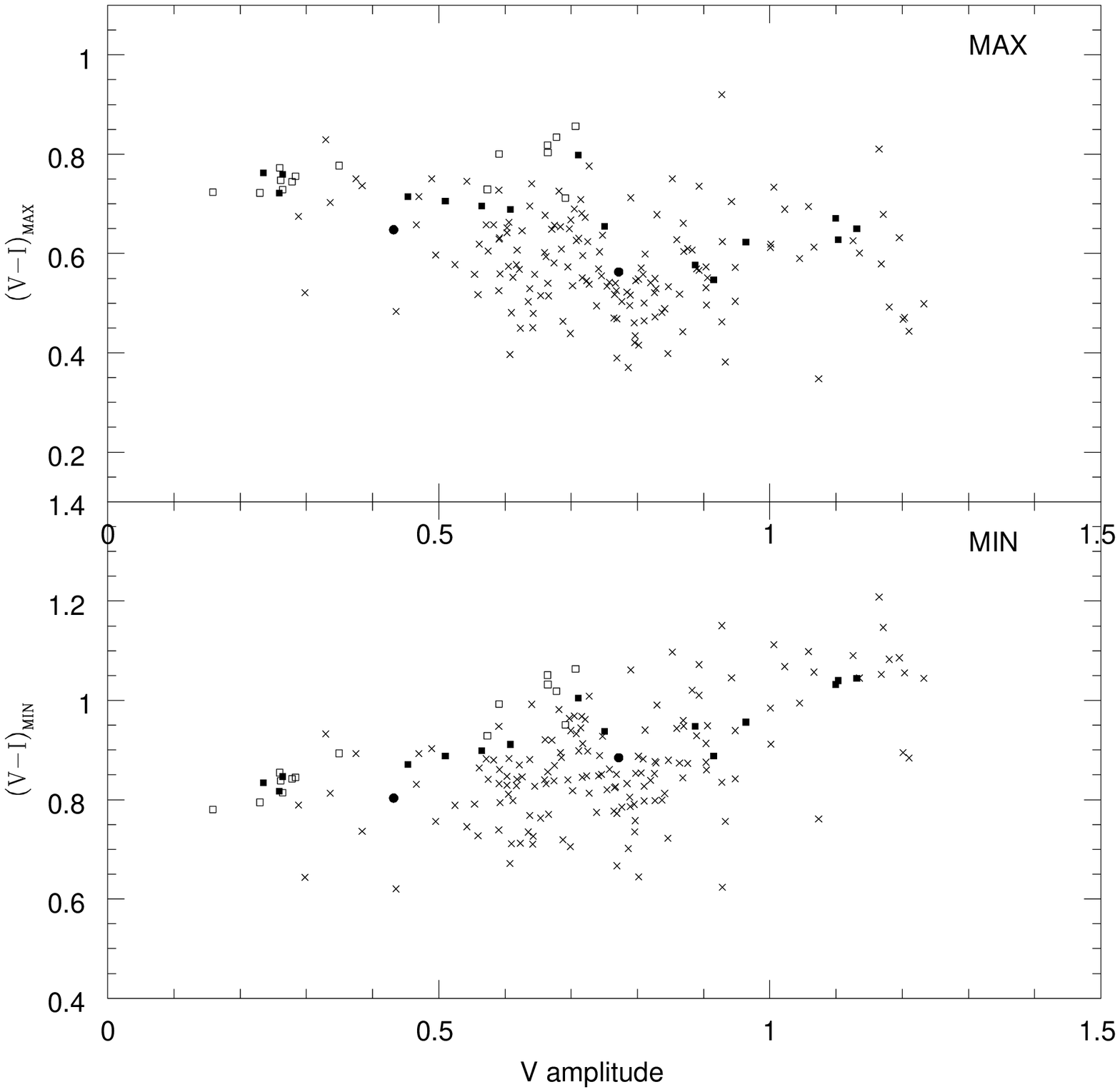}
    \epsfxsize=8.5cm \epsfbox{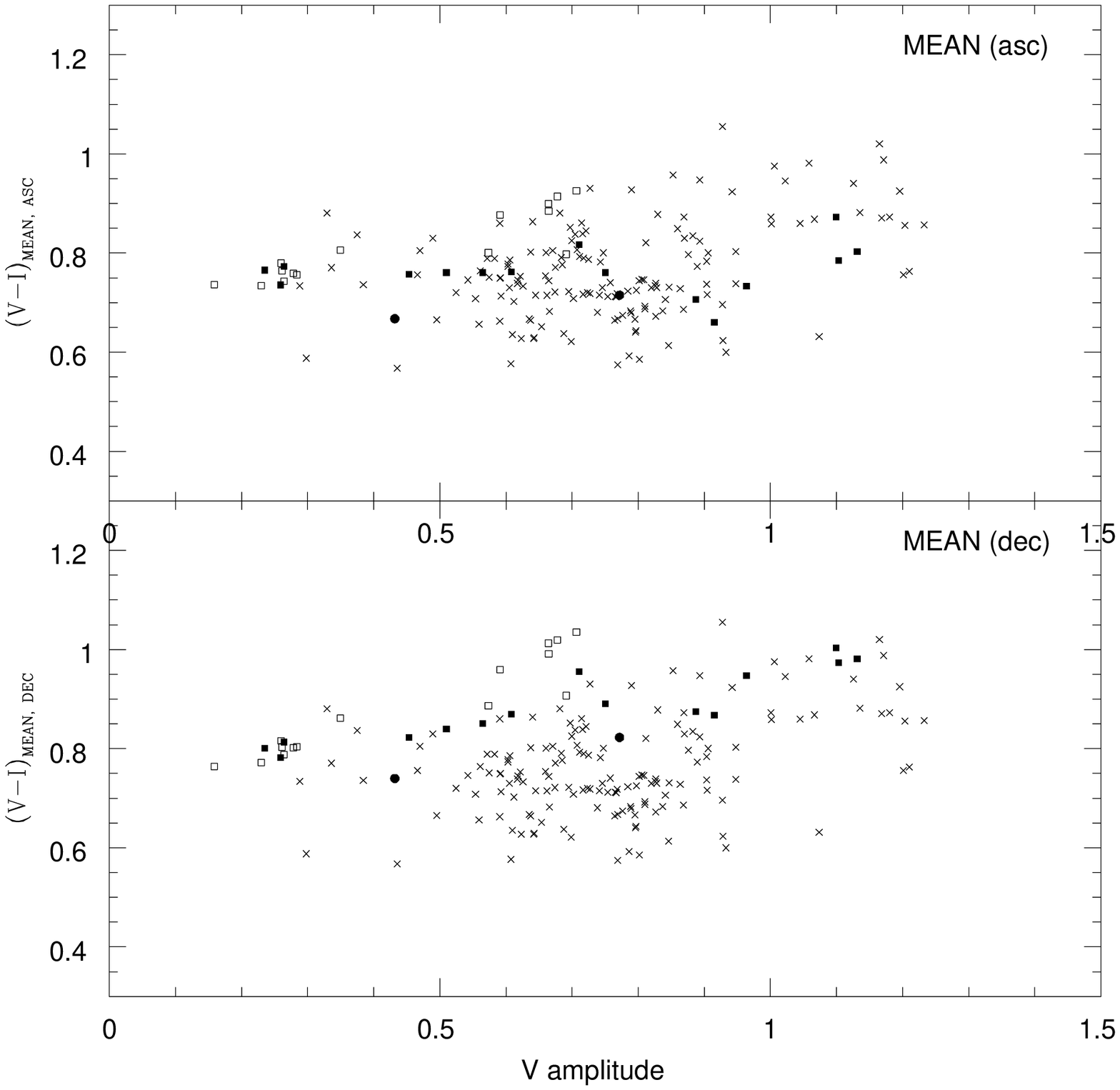}}
  \vspace{0cm}
  \caption{The V band amplitude-colour (AC) relations for the Galactic Cepheids. The symbols are same as in Figure \ref{fig1}. The bolometric model light curves are converted to V band light curves with the bolometric corrections obtained from BaSeL database. {\it Left} (a): AC relations at maximum and minimum light; {\it Right} (b): AC relation at mean light, for both of the acceding and descending branches. \label{fig11}.}
\end{figure*}

\section{Results}

     The results from the model calculations are summarized in Table \ref{tab1}, \ref{tab2} and \ref{tab3}. Table \ref{tab1} gives the input parameters, as well as the computed periods and growth rates from the linear non-adiabatic analysis for the fundamental and first overtone modes for the models constructed in this paper. We note that the models used generally had reasonably high fundamental mode growth rates and negative first overtone growth rates. This puts our models well inside the fundamental mode instability strip for all periods considered. Table \ref{tab2} gives the temperatures at the maximum/minimum luminosity at full amplitude for the corresponding models in Table \ref{tab1}. Table \ref{tab3} gives similar quantities but for the ascending and descending branch means. In Table \ref{tab3}, $<L>$ is the average luminosity obtained from the light curve in one pulsating cycle. $L_{mean}$ is the luminosity that is obtained from the closest temporal grid points (from the light curve) to the $<L>$, on both of the ascending and descending branches. $T_{mean}$ are the corresponding temperatures at these grid points. Since $<L>$ and $L_{mean}$ may not be equal to each other, we obtain the temperatures that correspond to $<L>$ by interpolation between $L_{mean}$ and the adjacent luminosity that bracket the $<L>$. These temperatures are referred as $T_{mean}^{inter}$ in Table \ref{tab3}. Note that the temperatures at mean light on the descending branch is normally cooler than the mean light temperatures on the ascending branch.



\begin{table*}
  \centering
  \caption{Temperatures at mean light from full-amplitude non-linear model calculations. See text for the meanings of $<L>$, $L_{mean}$, $T_{mean}$ and $T^{inter}_{mean}$. The periods, luminosity and temperature are in days, $L_{\odot}$ and $K$, respectively.}
  \label{tab3}
  \begin{tabular}{cccccccc} \hline
    $P$ & $<L>$ & $L_{mean}(asc)$ & $T_{mean} (asc)$ & $L_{mean} (des)$ & $T_{mean} (des)$ & $T_{mean}^{inter}$ (asc) & $T_{mean}^{inter}$ (des)\\ 
    \hline \hline
    \multicolumn{8}{c}{ML Relation from \citet{bon00}}
    \\ \hline
    43.3924    & 19774.9 & 19653.607 & 5076.72 & 19774.061 & 4719.16 & 5085.42 & 4719.24 \\
    37.8090    & 17280.2 & 17078.646 & 5114.30 & 17267.319 & 4759.22 & 5130.82 & 4760.52 \\
    34.4156    & 15390.5 & 15555.296 & 5159.32 & 15346.989 & 4772.37 & 5143.81 & 4776.56 \\
    30.8189    & 14018.4 & 14150.570 & 5205.60 & 14024.699 & 4833.74 & 5191.91 & 4833.08 \\
    23.4835    & 10192.1 & 10183.759 & 5239.51 & 10191.410 & 4931.29 & 5240.69 & 4931.39 \\
    20.0414    & 9841.01 & 9929.1677 & 5464.55 & 9828.9718 & 5093.87 & 5451.69 & 5095.78 \\
    15.8313    & 7180.69 & 7186.0066 & 5454.37 & 7176.7521 & 5161.14 & 5453.34 & 5162.00 \\
    12.1076    & 4954.75 & 4958.1150 & 5429.69 & 4955.1704 & 5254.78 & 5428.68 & 5254.65 \\
    10.0124    & 4147.76 & 4153.0371 & 5518.32 & 4150.1024 & 5390.79 & 5515.96 & 5389.95 \\
    9.69213    & 4150.28 & 4159.2783 & 5580.60 & 4141.0319 & 5433.01 & 5576.70 & 5436.31 \\
    8.83394    & 3600.06 & 3608.3159 & 5575.65 & 3611.7288 & 5439.77 & 5571.56 & 5435.18 \\
    8.31328    & 3270.50 & 3261.5882 & 5558.95 & 3271.9206 & 5435.65 & 5563.73 & 5435.04 \\
    7.12090    & 2738.62 & 2734.9282 & 5623.67 & 2739.4820 & 5492.58 & 5625.95 & 5492.13 \\
    6.16131    & 2271.74 & 2265.2783 & 5651.82 & 2275.3634 & 5533.36 & 5660.78 & 5531.01 \\
    5.59661    & 2149.59 & 2149.7132 & 5646.80 & 2151.3451 & 5559.70 & 5646.70 & 5558.46 \\
    \hline 
    \multicolumn{8}{c}{ML Relation from \citet{chi89}}
    \\ \hline 
    66.9094  & 34954.3 & 35127.180 & 5412.97 & 34950.980 & 4992.63 & 5404.79 & 4992.77 \\
    44.8456  & 19688.7 & 19843.004 & 5446.15 & 19660.664 & 4902.60 & 5434.34 & 4904.75 \\ 
    38.5673  & 15650.0 & 15888.680 & 5439.08 & 15652.258 & 4867.57 & 5417.09 & 4867.35 \\
    32.4902  & 11522.9 & 11331.742 & 5312.26 & 11511.286 & 4796.29 & 5335.61 & 4797.57 \\
    26.1109  & 10436.0 & 10650.701 & 5571.95 & 10432.457 & 4983.82 & 5541.43 & 4984.29 \\ 
    21.8715  & 9979.12 & 10039.792 & 5765.91 & 9909.8995 & 5244.96 & 5756.71 & 5258.68 \\
    16.9148  & 6648.78 & 6722.9877 & 5704.33 & 6649.7902 & 5213.82 & 5688.05 & 5213.56 \\
    12.4229  & 3945.54 & 3988.1625 & 5571.12 & 3927.3390 & 5146.88 & 5555.92 & 5153.68 \\
    10.0049  & 3007.79 & 3028.3688 & 5566.36 & 3004.3297 & 5237.81 & 5557.55 & 5239.67 \\
    9.37928  & 2802.28 & 2810.9339 & 5570.03 & 2810.0216 & 5283.35 & 5566.27 & 5278.95 \\
    8.80955  & 2605.42 & 2606.5476 & 5570.03 & 2609.8551 & 5319.08 & 5569.55 & 5316.25 \\
    8.21323  & 2417.35 & 2413.1797 & 5579.92 & 2422.3123 & 5369.50 & 5581.37 & 5366.15 \\
    7.06755  & 1914.78 & 1919.5566 & 5532.01 & 1908.3662 & 5401.51 & 5527.21 & 5406.84 \\
    5.56235  & 1764.77 & 1759.8178 & 5554.42 & 1763.3763 & 5440.10 & 5559.56 & 5441.30 \\
    5.60322  & 1490.55 & 1495.4280 & 5647.41 & 1495.1755 & 5507.79 & 5641.55 & 5503.47 \\
    \hline 
    \multicolumn{8}{c}{Input parameters from SKM}
    \\ \hline
    10.5472   & 2022.70 & 2028.8310 & 5837.90 & 2031.3158 & 5634.15 & 5832.38 & 5627.78 \\
    6.25949   & 3777.86 & 3731.9089 & 5710.18 & 3785.7844 & 5373.31 & 5726.55 & 5369.91 \\
    \hline 
  \end{tabular}
\end{table*}

     
\subsection{The PC relations}

     In this subsection we compare the PC relations from the model calculations and the observations. Figure \ref{fig1} shows a four panel plot of log-period against extinction corrected $(V-I)$ colour for Galactic Cepheids as given in KN. Superimposed on this are values obtained from our models, where the theoretical temperatures are converted to $(V-I)$ colours using method (ii) described above. The solid and open squares represent models based on the \citet{chi89} and \citet{bon00} ML relations respectively, whereas the solid circles refer to models computed with the ML relation used in SKM. The left panels are for maximum (top) and minimum (bottom) light respectively, whilst the right panels are for the means of the ascending and descending branches, respectively. We see that there is very good agreement between models and observations in this diagram. Figure \ref{fig2} shows the same quantities but on the $\log(P)$-$\log(T)$ plane. The observed $(V-I)$ colours for the Galactic Cepheid data are converted to the temperature by using the BaSeL database mentioned in previous section, where the $\log (g)$ terms for the observed colours are approximated by using $\log (g)=2.62-1.21\log(P)$ \citep{bea01}. We note that the qualitative nature of our results is the same in Figure \ref{fig1} and \ref{fig2}: the flat nature of the relation at maximum light, the non-zero slope at minimum light and an amalgam of these at mean light. 

     Recall that KN analyzed a large sample of Galactic Cepheid data and showed that the PC relation was consistent with a single line at mean and minimum light, but at maximum light there was significant statistical evidence for a break at 10 days. Figures \ref{fig1} \& \ref{fig2} suggests that this break actually occurs at $\log (P) \approx 0.8$. As in KN, we can perform a two line regression, for Cepheids with periods smaller and greater than $\log (P) = 0.8$, and compare with a one line regression across the entire period range. This is what is meant by short and long period in the following paragraph. Table \ref{tab4} shows the phase, overall dispersion, slope and error at that phase and then these same quantities for the short and long period Cepheids respectively. From Figure \ref{fig2} and Table \ref{tab4}, we can conclude the following results:


\begin{table*}
  \centering
  \caption{$\log (T)$ vs $\log (P)$ regression at maximum, mean and minimum light for the Galactic Cepheid data. $\sigma$ is the dispersion from the regression of the form: $\log(T)=a+b\log(P)$. The subscripts $_L$ and $_S$ refer to long ($\log(P)>0.8$) and short period Cepheids, respectively.}
  \label{tab4}
  \begin{tabular}{lccccccccc} \hline
    phase & $\sigma_{all}$ & $b_{all}$ & $a_{all}$ & $\sigma_L$ & $b_L$ & $a_L$ & $\sigma_S$ & $b_S$ & $a_S$ \\
    \hline \hline
    Max  & 0.016  & $-0.033\pm0.005$ & $3.807\pm0.005$ & 0.017 & $-0.014\pm0.009$ & $3.787\pm0.009$ & 0.014 & $-0.109\pm0.023$ & $3.862\pm0.016$ \\
    
    Mean & 0.017  & $-0.062\pm0.006$ & $3.801\pm0.005$ & 0.019 & $-0.068\pm0.010$ & $3.808\pm0.010$ & 0.014 & $-0.131\pm0.023$ & $3.846\pm0.016$ \\
    Min  & 0.022  & $-0.077\pm0.007$ & $3.786\pm0.007$ & 0.023 & $-0.076\pm0.012$ & $3.786\pm0.012$ & 0.019 & $-0.193\pm0.031$ & $3.865\pm0.022$ \\
    \hline
  \end{tabular}
\end{table*}

     \begin{enumerate}
     \item  The overall slope in the $\log (P)$-$\log (T)$ plane at maximum light is flat and is significantly different from the slope at mean and minimum light. 
       
     \item  The short period slopes in this same plane are significantly different from zero at all three phases. In contrast, the long period slope at maximum light in this plane is close to zero and is significantly different from the slope at mean and minimum light.

     \item  If we reject all Cepheids with $\log (P) < 0.8$ and test, using the methodology of KN, whether the remaining data are consistent with a single line or with two lines broken at a period of 10 days, we find that in the $\log (P)$-$\log (T)$ plane, even at maximum light, the data are consistent with a single line. Thus the broken PC relation at maximum light for Galactic Cepheids described by KN is due primarily to the different PC relation obeyed by short ($\log (P)<0.8$) period, fundamental mode Cepheids.
     \end{enumerate}

     We note that this good agreement exists for maximum, mean and minimum light. The exception is perhaps the longer period models constructed with the \citet{bon00} ML relation, with $\log (P) > 1.3$, at maximum light. We discuss these models in Section 4. Since the observed data exhibit very similar characteristics in both period-color and period-temperature planes, that is, a distinctly non-linear relation with a sudden break at a period of $0.75 < \log(P)< 1.0$ and since the models agree well with the data, it follows that existing models contain the physics to explain these observations. We study this in Section 3.3

\subsection{The AC relations}

     Figure \ref{fig11} depicts plots of the AC relations for both models and data, at maximum, minimum, ascending and descending branch means. The bolometric corrections obtained from the BaSeL database are used to construct the V band amplitudes from the model light curves. We see good agreement between models and data in these diagrams, though models constructed with our code using the \citet{bon00} ML relation have smaller amplitudes even though these models match observations in the AC diagrams. Models constructed using the \citet{chi89} ML relation with our code generally have larger amplitudes.  We caution here that the computed amplitudes, unlike the periods, depend strongly on the strength of the artificial viscosity and of the assumed turbulent eddy viscosity.

\begin{figure*}
       \vspace{0cm}
       \hbox{\hspace{0.2cm}\epsfxsize=8.5cm \epsfbox{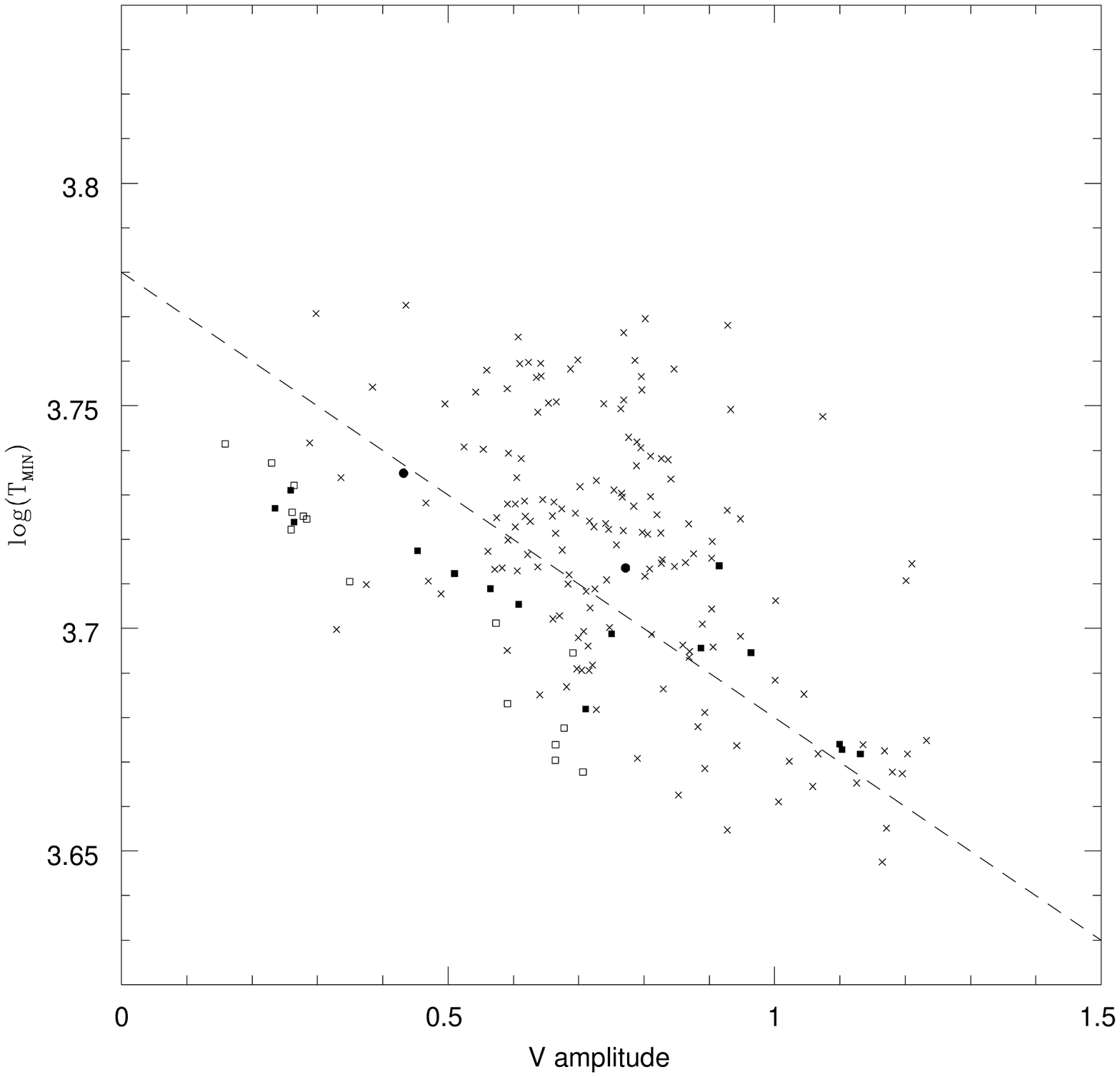}
         \epsfxsize=8.5cm \epsfbox{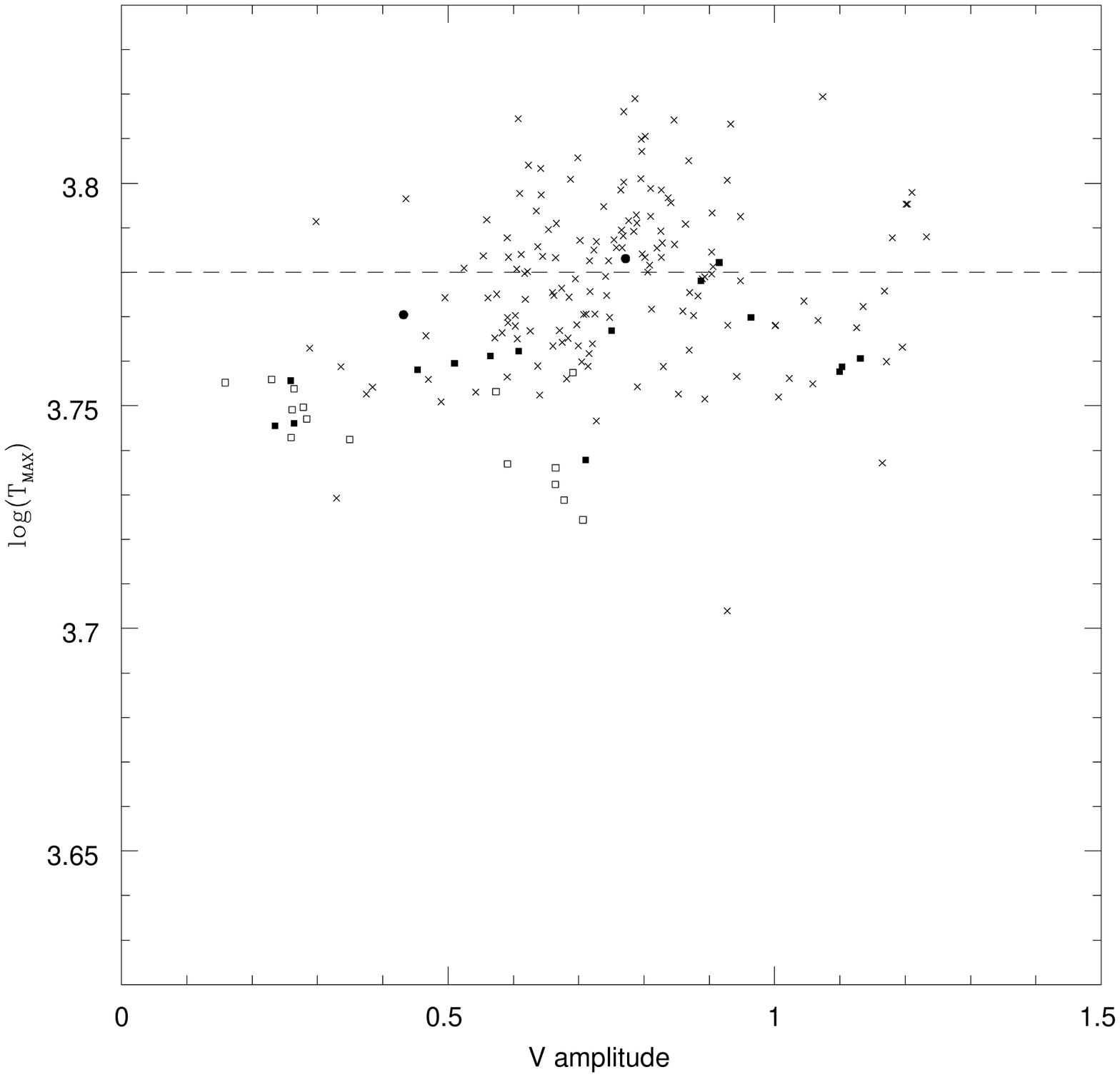}}
       \vspace{0cm}
       \caption{The plots of $\log(T)$ vs. $V$ amplitudes ($V_{amp}$) for the Galactic data and models. The symbols are same as in Figure \ref{fig1}. The conversion from observed $(V-I)$ colour to temperature is done with the BaSeL database (see text for details). {\it Left} (a): The plot of $V_{amp}$-$\log(T_{min})$, The dashed line is {\it not} the best-fit regression to the data, but illustrates how closely the overall trend of the data follow a slope of $-0.1$, as predicted from equation (1). This line is constructed with $\log (T_{min})=3.78-0.1V_{amp}$. {\it Right} (b): The plot of $V_{amp}$-$\log(T_{max})$. The dashed line show the average value of $\log (T_{max})$, where $<\log (T_{max})>=3.78$. \label{fig3}}
     \end{figure*}

     Table \ref{tab5} gives the results from a linear regression of $V$ band amplitude against $\log T$ at the three phases of interest for the Galactic Cepheid data, taken from Figure \ref{fig2} \& \ref{fig11}. Equation (1) shows that, if the variation of $\log T_{max}$ with period has a shallow slope close to zero, then an equation of the form $\log T_{min} = a + bV_{amp}$ is such that $b$ is about $-0.10$. From Table \ref{tab5}, we see that the overall slope is $-0.079$, which is very close to the theoretically expected value. Accounting for the error on the slope shows that the actual slope is statistically indistinguishable from the theoretical value of $-0.10$. In addition, Table \ref{tab5} implies that the overall slope for $V_{amp}$-$\log T_{max}$ relation is close to zero across the entire period range. This strongly supports the validity of equation (1) and our interpretation of the predictions arising from this equation if either $\log T_{max}$ or $\log T_{min}$ obey a flat relation with period. Figure \ref{fig3} argues this graphically with a two panel plot displaying $V$ band amplitude against $\log T_{min}$ and $\log T_{max}$ on the left and right panels, respectively. An AC relation may also occur if either $\log T_{max}$ or $\log T_{min}$ has a shallower relation with period than the other quantity.


\begin{table*}
  \centering
  \caption{V band amplitude ($V_{amp}$) vs. $\log T_{min}$ regression for the Galactic Cepheid data. $\sigma$ is the dispersion from the regression of the form: $\log T_{min}=a+bV_{amp}$. The subscripts $_L$ and $_S$ refer to long ($\log(P)>0.8$) and short period Cepheids, respectively.}
  \label{tab5}
  \begin{tabular}{lcccccccccc} \hline
    phase & $\sigma_{all}$ & $b_{all}$ & $a_{all}$ & $\sigma_L$ & $b_L$ & $a_L$ & $\sigma_S$ & $b_S$ & $a_S$ \\
    \hline \hline
    Max   & 0.018 & $ 0.011\pm0.007$ & $3.770\pm0.006$ & 0.017 & $ 0.018\pm0.008$ & $3.757\pm0.007$ & 0.015 & $ 0.038\pm0.015$ & $3.760\pm0.011$ \\
    Mean  & 0.021 & $-0.050\pm0.009$ & $3.784\pm0.007$ & 0.021 & $-0.048\pm0.009$ & $3.776\pm0.009$ & 0.017 & $-0.007\pm0.017$ & $3.761\pm0.012$ \\
    Min   & 0.024 & $-0.079\pm0.010$ & $3.778\pm0.008$ & 0.023 & $-0.073\pm0.011$ & $3.766\pm0.009$ & 0.023 & $-0.040\pm0.023$ & $3.760\pm0.016$ \\
    \hline
  \end{tabular}
\end{table*}

\subsection{The interaction of the HIF and photosphere}

\begin{figure*}
       \vspace{0cm}
       \hbox{\hspace{0.2cm}\epsfxsize=8.5cm \epsfbox{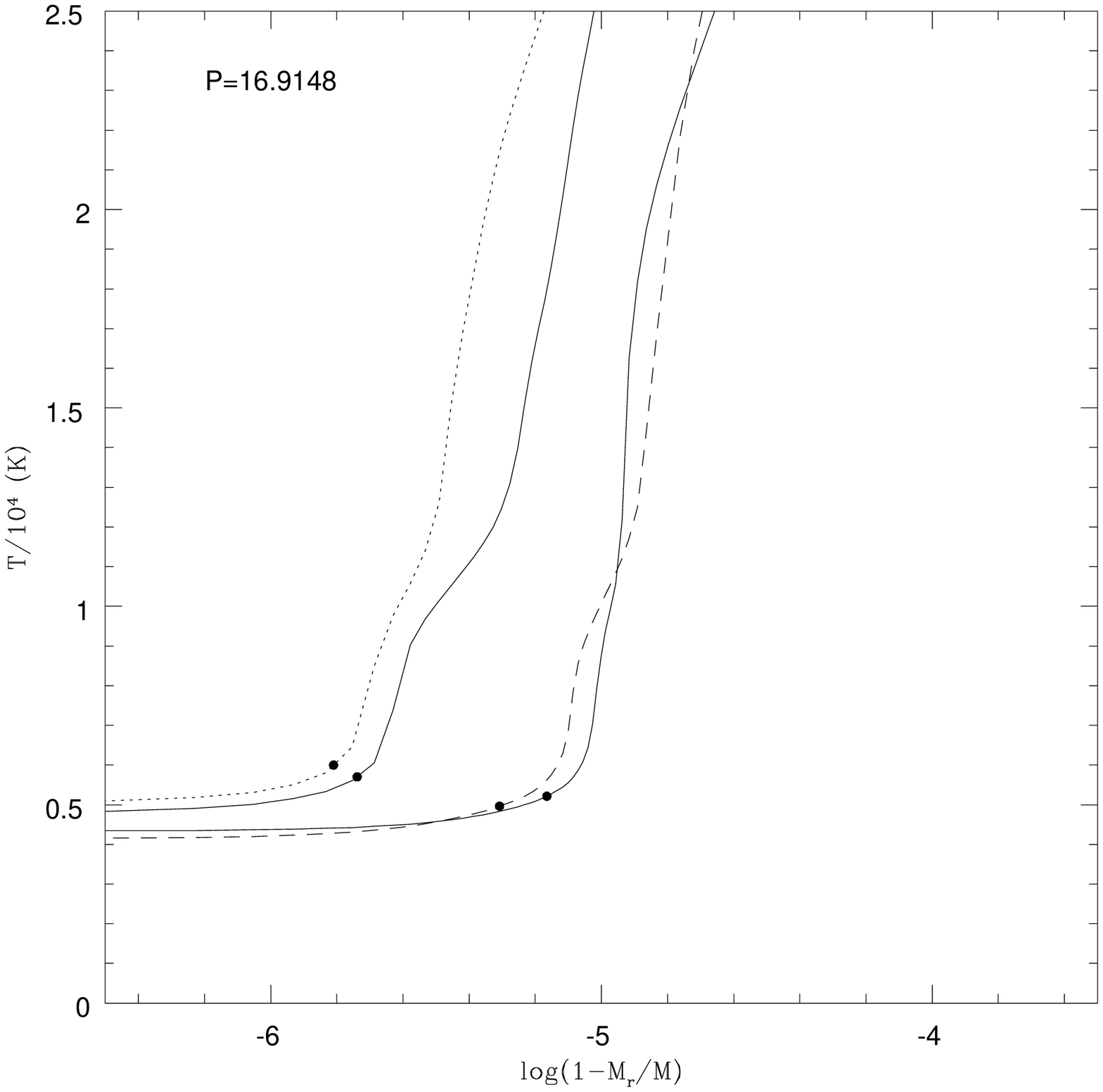}
         \epsfxsize=8.5cm \epsfbox{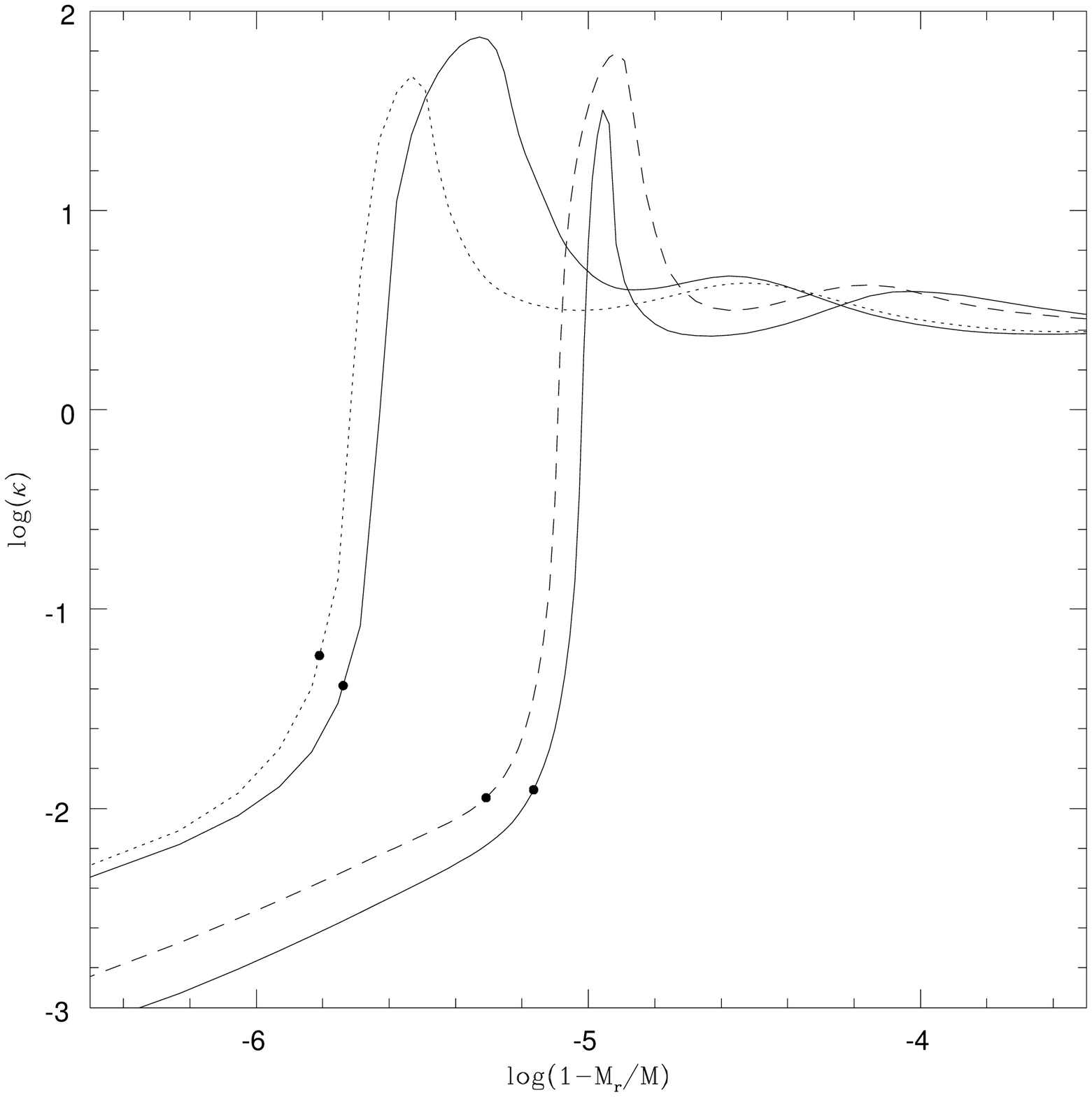}}
       \vspace{0cm}
       \caption{The temperature and opacity profiles for a long period model. The dotted, solid and dashed curves are for the profiles at maximum, mean and minimum light, respectively. The filled circles are the location of the photosphere at $\tau=2/3$ for each phases. The mean light profiles at the ascending and descending branch are the solid curves that lie close to the profiles at maximum light (dotted curves) and minimum light (dashed curves), respectively. {\it Left} (a): Temperature profile; {\it Right} (b): Opacity profile. \label{fig4}}
     \end{figure*}

\begin{figure*}
       \vspace{0cm}
       \hbox{\hspace{0.2cm}\epsfxsize=8.5cm \epsfbox{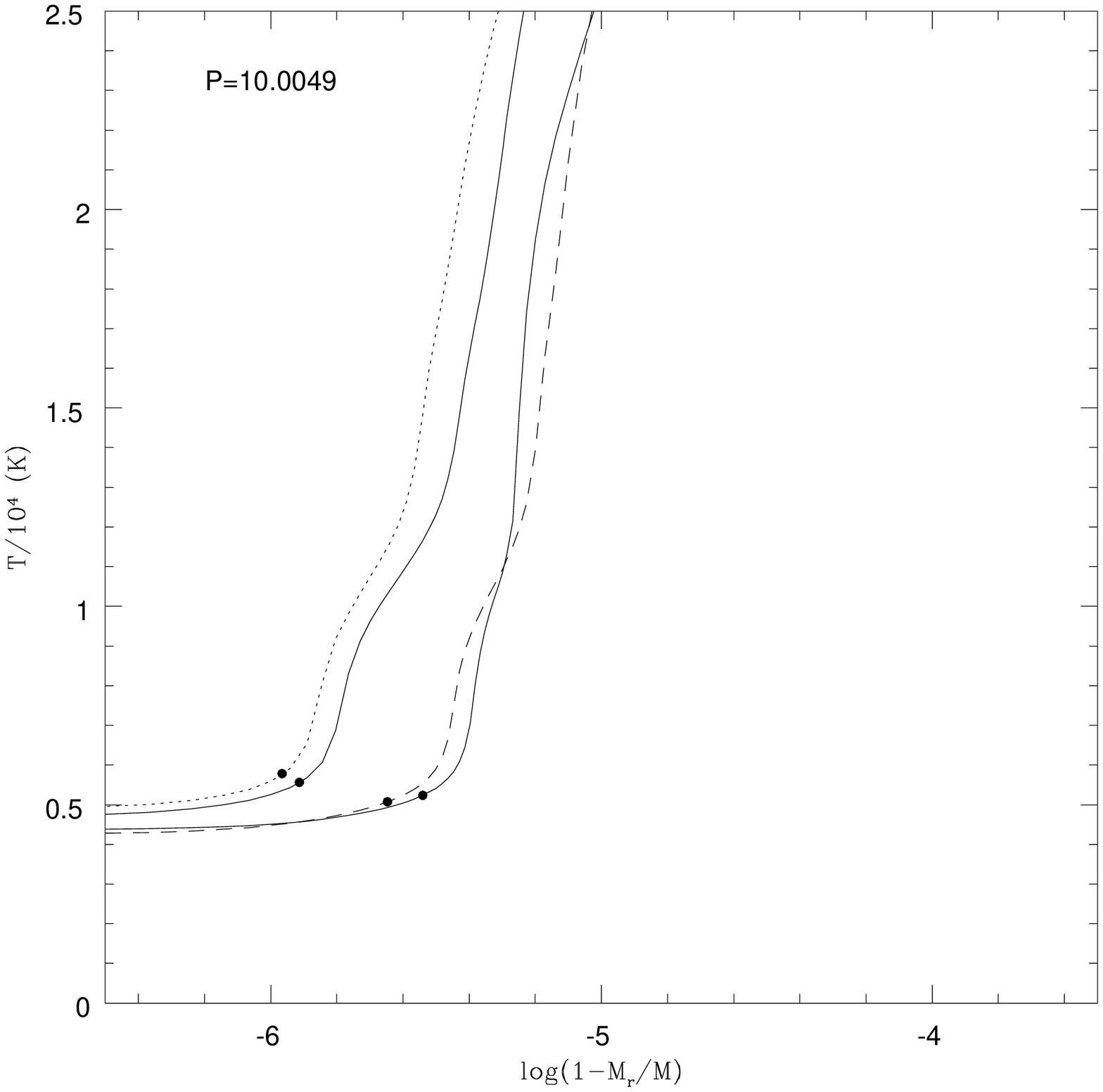}
         \epsfxsize=8.5cm \epsfbox{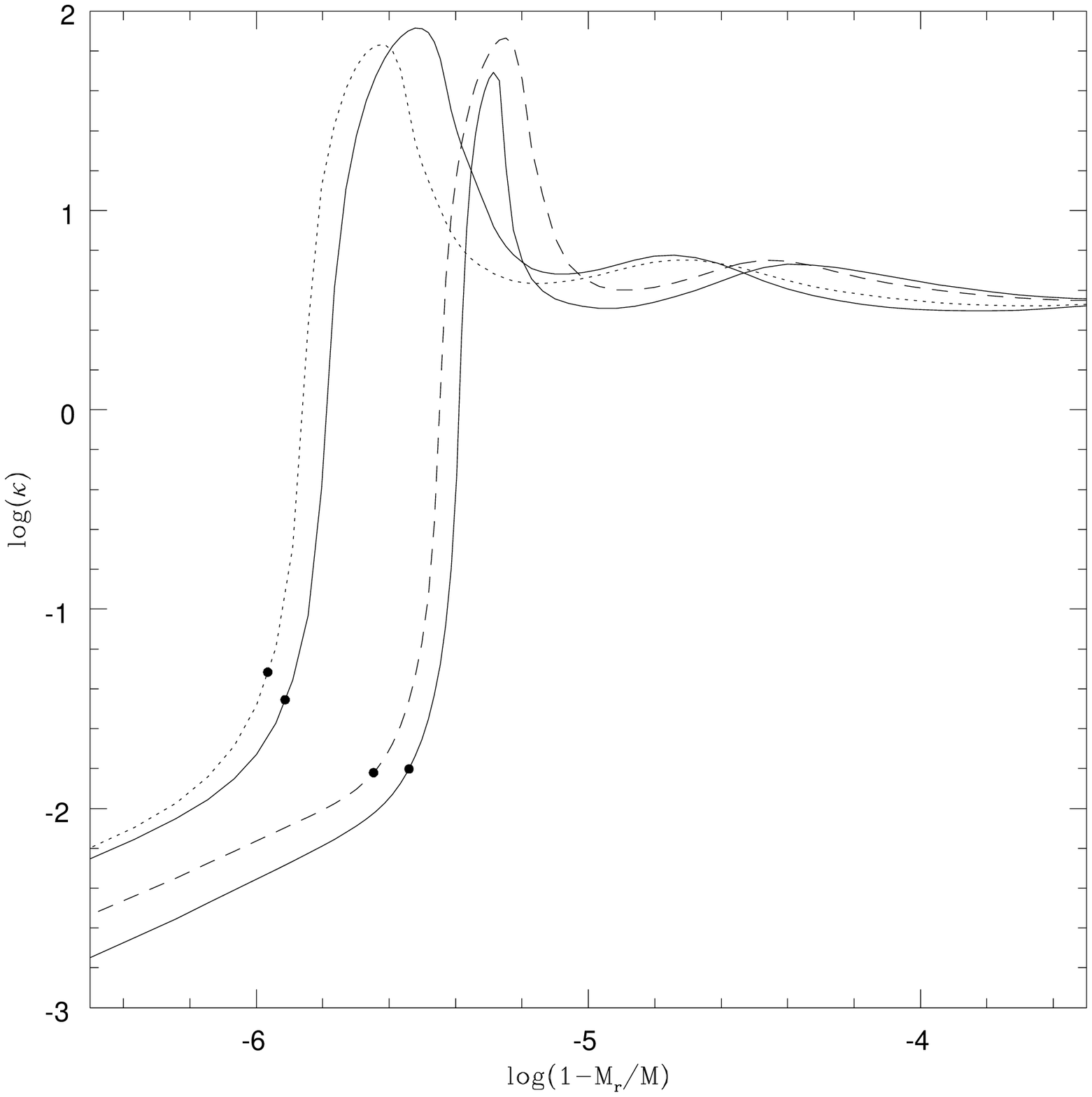}}
       \vspace{0cm}
       \caption{Same as Figure \ref{fig4}, but for model with period at 10 days. \label{fig5}}
     \end{figure*}

\begin{figure*}
       \vspace{0cm}
       \hbox{\hspace{0.2cm}\epsfxsize=8.5cm \epsfbox{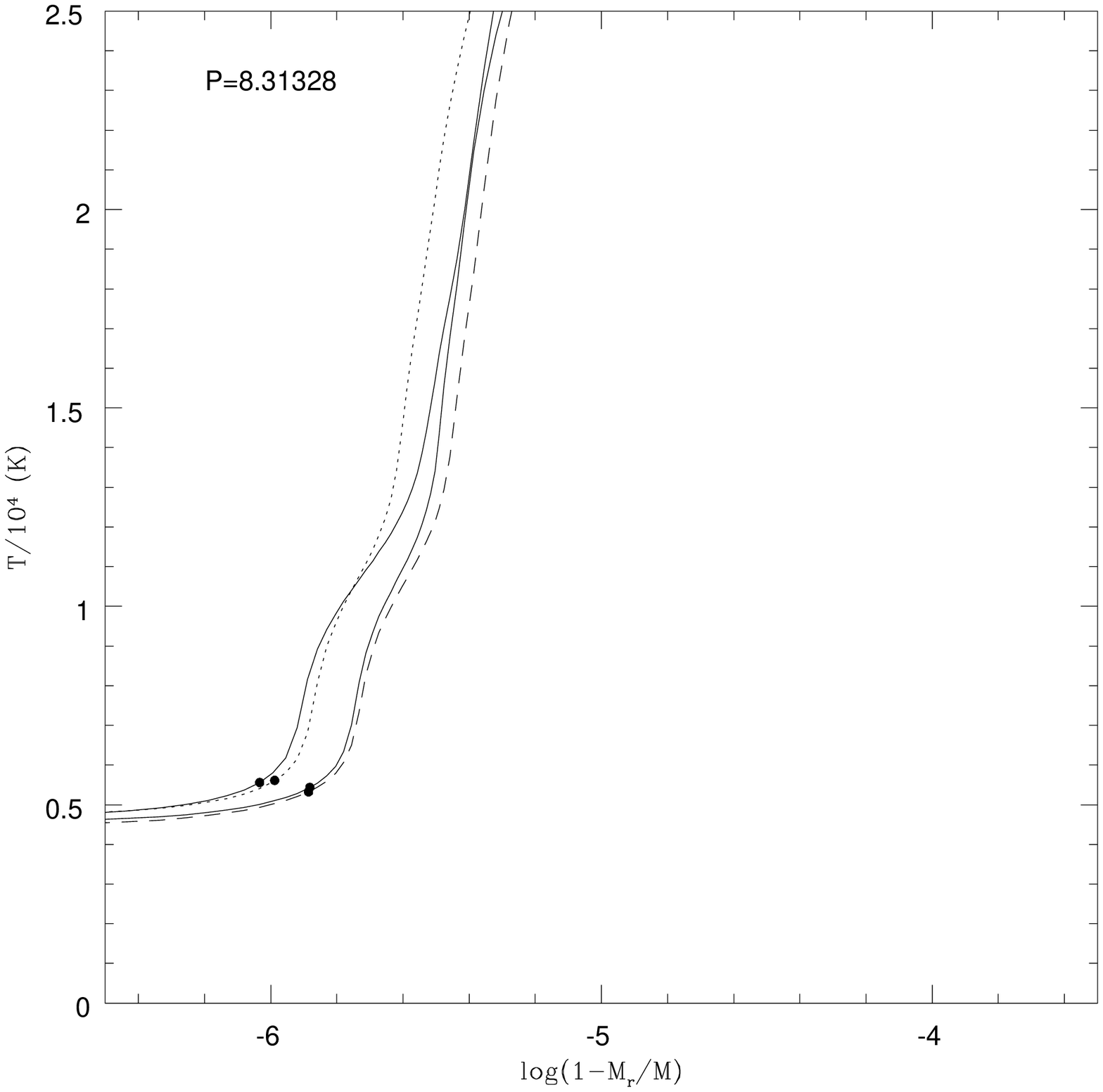}
         \epsfxsize=8.5cm \epsfbox{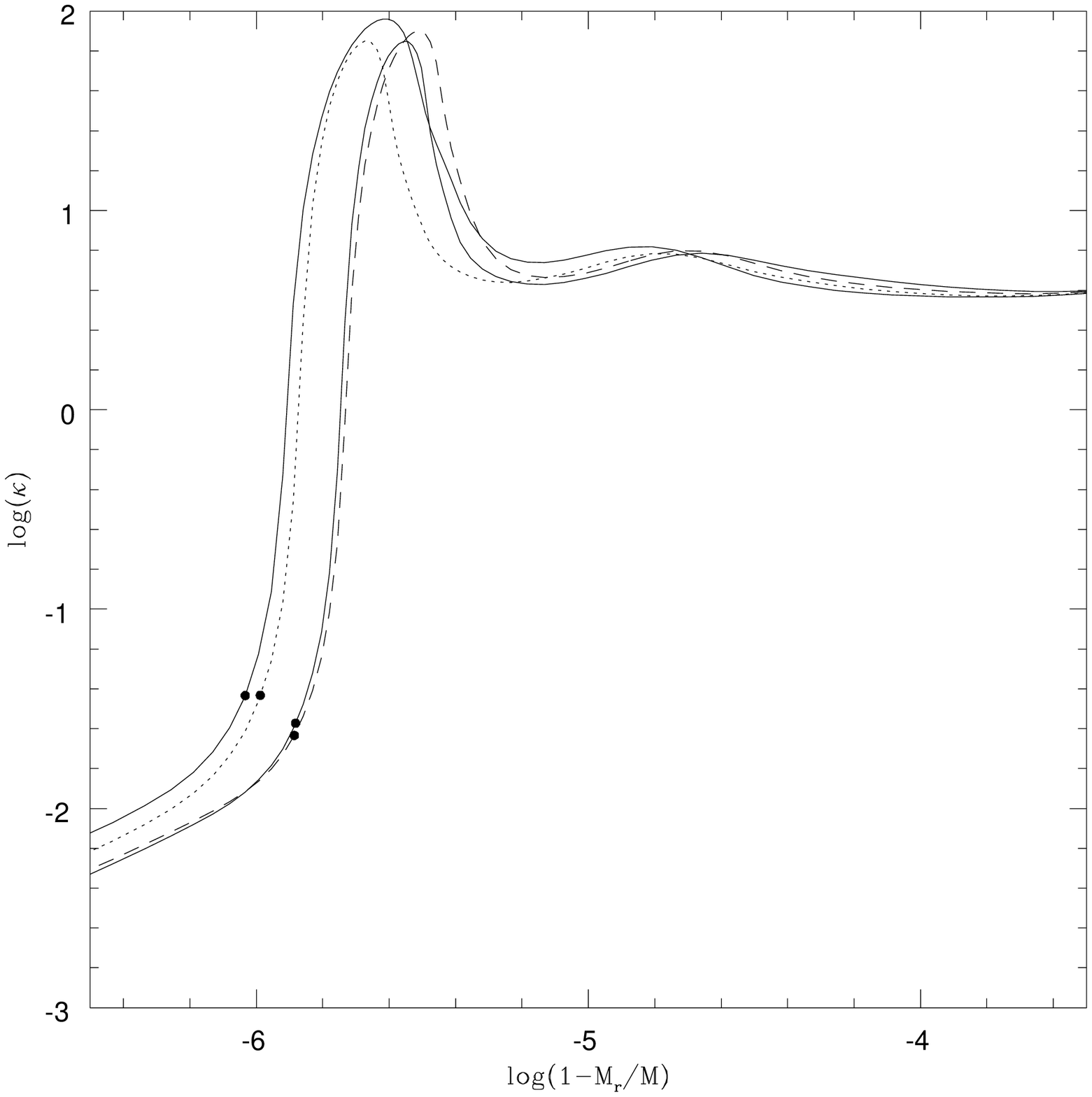}}
       \vspace{0cm}
       \caption{Same as Figure \ref{fig4}, but for model with short period. \label{fig6}}
     \end{figure*}

\begin{figure*}
       \vspace{0cm}
       \hbox{\hspace{0.2cm}\epsfxsize=8.5cm \epsfbox{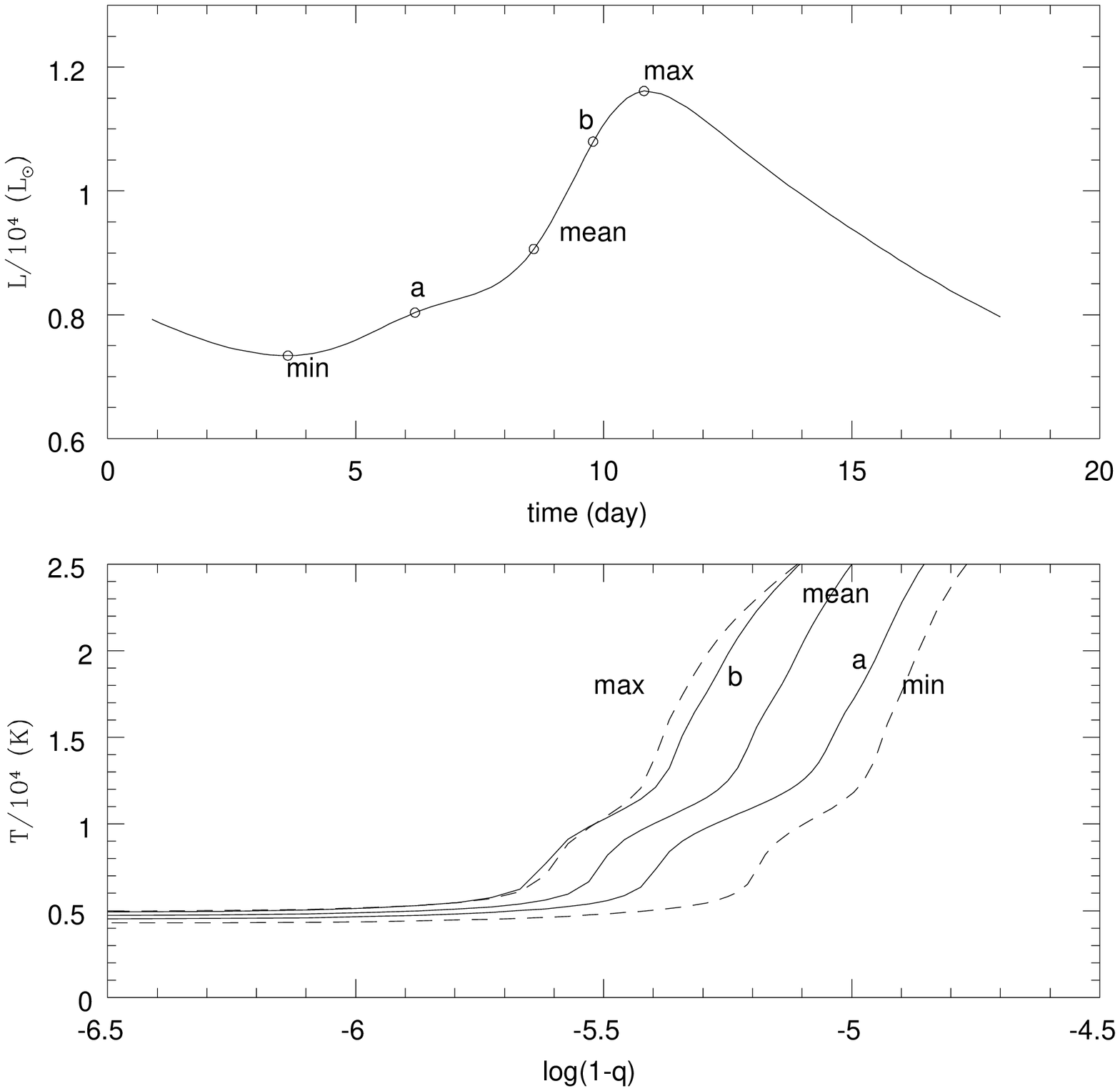}
         \epsfxsize=8.5cm \epsfbox{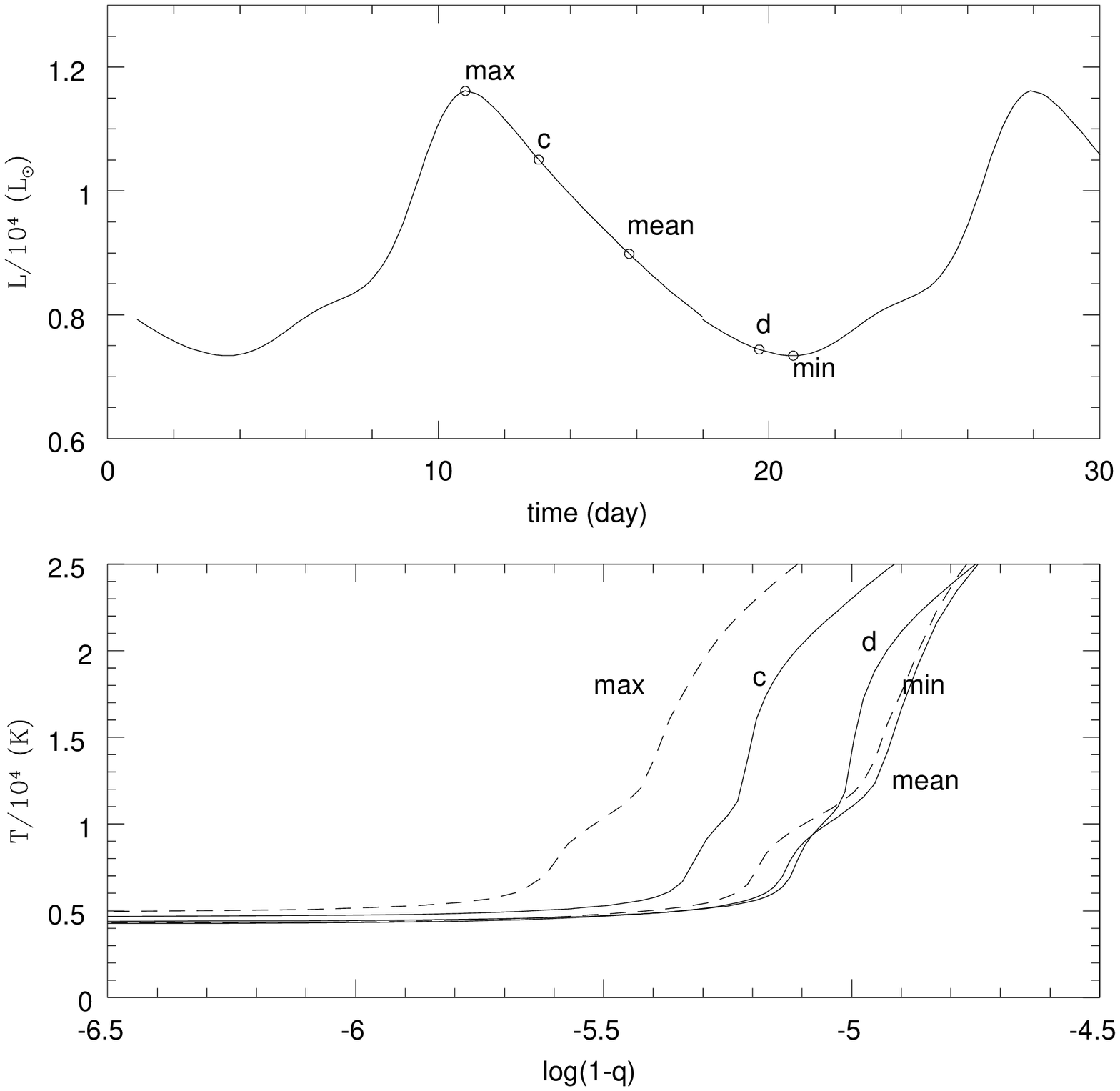}}
       \vspace{0cm}
       \caption{Illustrations of the temperature profiles at different pulsation phases, where $q=M_r/M$. The upper and lower panels are the light curves and the temperature profiles at various phases, respectively. The lower-left panel shows the changes of temperature profiles from minimum light to maximum light (ascending branch), and the lower-right panel shows the similar temperature profiles in descending branch. In lower panels, the dashed curves represent the temperature profiles at maximum and minimum light, while the solid curves represent the temperature profiles at other phases as indicated in upper panels. \label{fig7}}
     \end{figure*}

     In Figures \ref{fig4}-\ref{fig6}, we present the plots of the temperature and opacity profiles, i.e. temperature and log opacity against location in the mass distribution respectively, for a representative selection of the models in Table \ref{tab1}. We display plots for a long period ($\log (P) > 1.0$), short period ($\log (P) < 1.0$) and a 10 day period model. The mass distribution is measured by the quantity $Q = \log (1-M_r/M)$, where $M_r$ is the mass within radius $r$ and $M$ is the total mass. Each panel shows the temperature and opacity profiles at maximum, mean and minimum light with dotted, solid and dashed curves, respectively. For each model, we also include the plots for the case when the mean is on the ascending and descending branch respectively. The photosphere is marked with a filled circle in these figures. Figure \ref{fig7} illustrates the luminosity and temperature profiles against mass distribution in one pulsating cycle.


\begin{figure}
       \hbox{\hspace{0.1cm}\epsfxsize=7.5cm \epsfbox{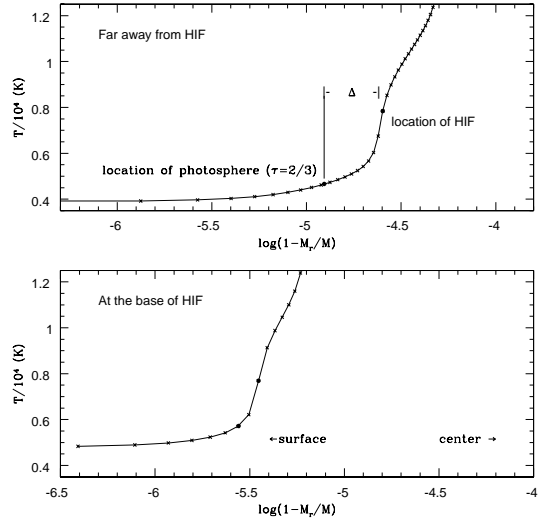}}
       \caption{Illustration of the location of HIF and photosphere, and the way to calculate $\Delta$. The crosses represent the zones in the model calculations. The location of the photosphere, at $\tau=2/3$, and the HIF are marked with filled circles. The location of the HIF is defined as the zone with the steepest gradient in the temperature profile. Therefore, $\Delta$ is the ``distance'' (in terms of $\log[1-M_r/M]$) between the photosphere and HIF. Upper and lower panels illustrate the cases that the photosphere is far away and close to the HIF, respectively.}
       \label{fig8}
     \end{figure}

     In general, the HIF sweeps back and forth in the mass distribution as the star pulsates. A naive way of thinking about this would be that the HIF is furthest out and furthest in, in the mass distribution, at maximum and minimum light respectively (SKM). Our models indicate that the situation is more complicated than this, and we summarize the main results from Figure \ref{fig4}-\ref{fig6} as follow:

     \begin{enumerate}
     \item  At minimum light, the photosphere is closest to the HIF at short period but moves progressively away from the base of the HIF as the model period increases. This is in concordance with the lowest panel of Figures \ref{fig1} and \ref{fig2} which display the period temperature relation for models and data as one single line such that longer period models have cooler photospheric temperatures at minimum light. However, at maximum light, the photosphere lies at the base of the of the HIF for all the models presented.

     \item  We can quantify this by defining the quantity $\Delta$ to be the ``distance'', in units of $Q$, between the photosphere and the point in the HIF where $dT/dQ$ is a maximum. Figure \ref{fig8} illustrates this and Figure \ref{fig9} portrays the results when we plot $\Delta$ against $\log (P)$ at maximum, minimum and the ascending and descending branch means. In Figure \ref{fig9}, the error bars are estimated from the coarseness of the grid points around the location of HIF. We see clearly that at maximum light, $\Delta$ is constant over a large period range, whereas at minimum light, $\Delta$ gets larger as the period increases. The value of $\Delta$ at the phases of ascending and descending branch means is a mixture of the behavior at maximum and minimum light. There is an indication that for long period models ($\log (P) > 1.3$) at maximum light, the photosphere starts to become disengaged from the base of the HIF. As the period increases, so does the $L/M$ ratio and, in general, the effective temperatures become lower. Kanbur (1995) showed that such changes push the HIF further inside the mass distribution, making it harder to interact with the photosphere even at maximum light when it is closer to the surface. This can cause a non-zero slope in the $\log (P)$-$\log (T_{max})$ plane at long periods. This was also noted by SKM.
 
     \item  The HIF at mean light is not always between its locations at maximum or mean light. Figure \ref{fig7} illustrates our point and shows various locations on the ascending and descending branch of a typical light curve from our models. The lower two panels show the temperature profile at each of these phases. We see that the HIF at phase $b$ is actually further out in the mass distribution than at maximum light and conversely the HIF is further in in the mass distribution at the phase corresponding to mean light on the descending branch than at minimum light. Moreover, the nature and shape of the HIF seems to change significantly during the pulsation period.

     \item  At short/long period, the HIF at ascending/descending branch mean light is located in front/behind its location at maximum/minimum light.

     \item  We also note that the nature of the HIF changes when it is located outside of its locations at either maximum or minimum light. These changes in the HIF occur at period ranges corresponding to the location of bumps on the ascending or descending branches of the light curve: the Hertzsprung progression.

     \item  If we define "amplitude of HIF oscillation" as the range of $\log (1-M_r/M)$ values occupied by the HIF, then there is a correlation between this and the resulting amplitude of the photospheric light curve.
     \end{enumerate}

\subsection{Cepheids with $\log (P) < 0.8$}

\begin{figure*}
       \vspace{0cm}
       \hbox{\hspace{0.2cm}\epsfxsize=8.5cm \epsfbox{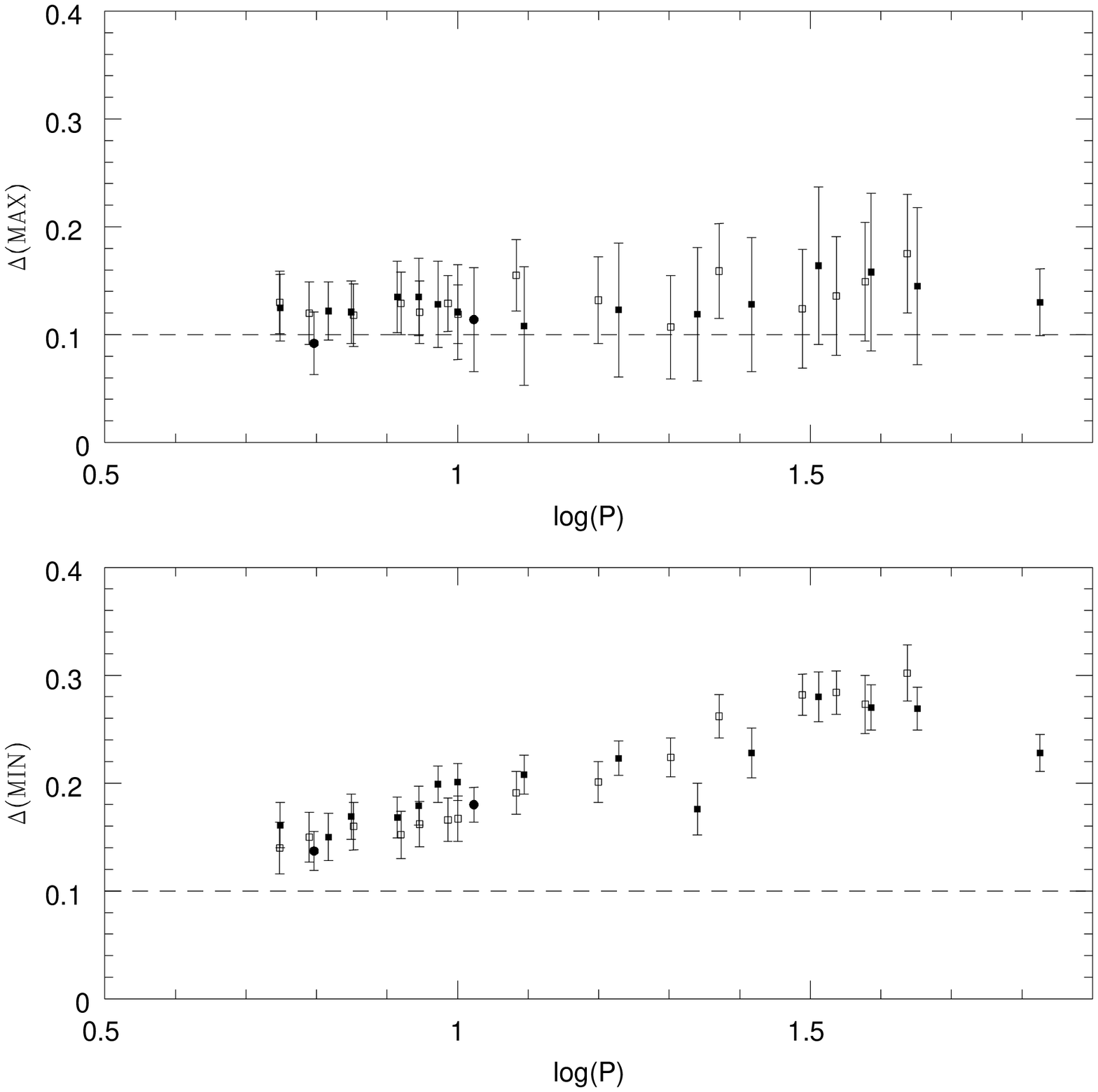}
         \epsfxsize=8.5cm \epsfbox{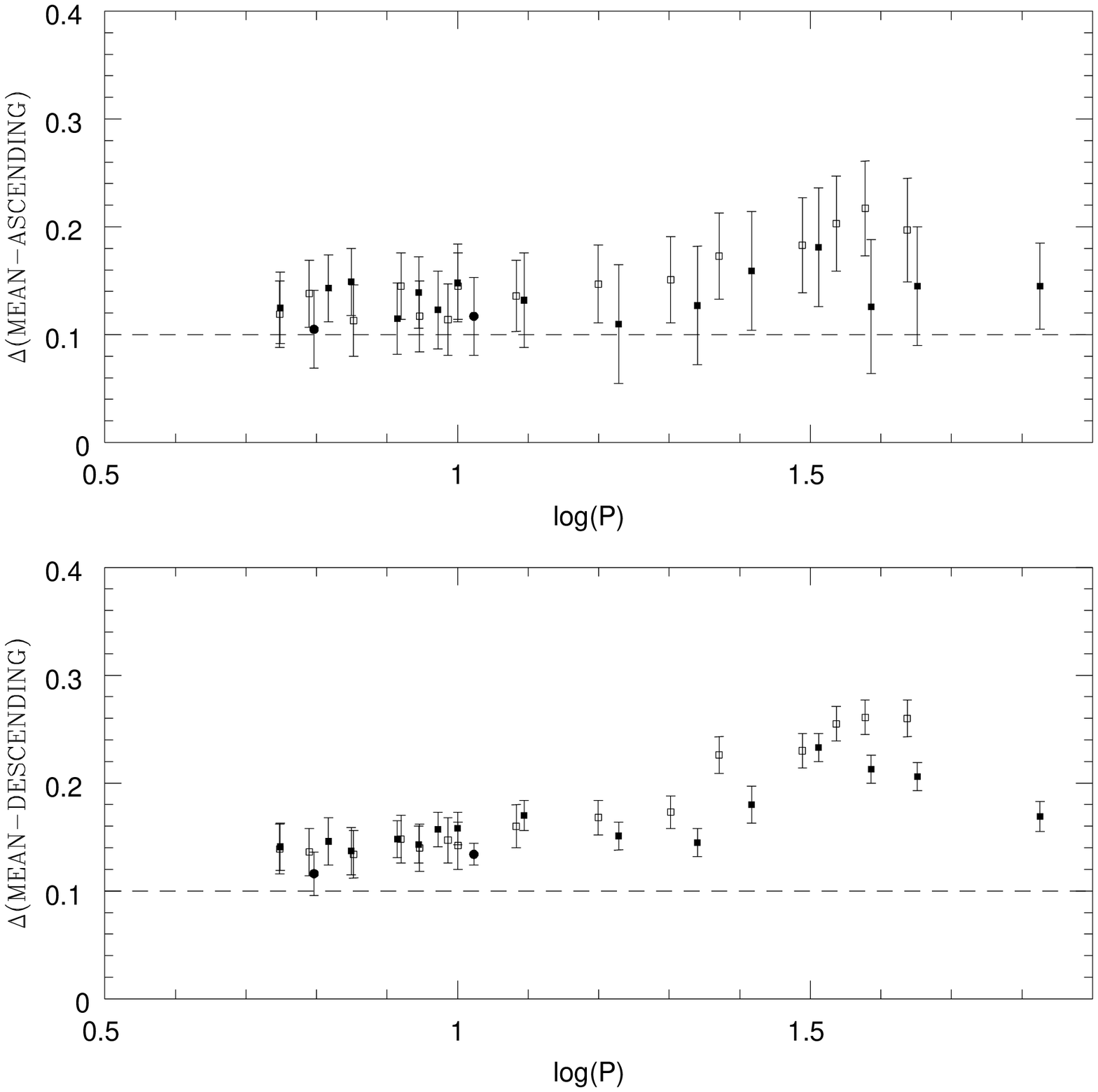}}
       \vspace{0cm}
       \caption{The plots of $\Delta$ as function of $\log(P)$. The symbols are same as in Figure \ref{fig1}. The dashed lines represent (roughly) the outer boundary of the HIF.  \label{fig9}}
     \end{figure*}

     Figure \ref{fig1} and the results of Tables \ref{tab4} \& \ref{tab5} suggest that Cepheids with $\log (P) < 0.8$ do not obey a flat relation with period at maximum light. Our current grid of models does not include any models with such periods, but Figures \ref{fig9} does indicate that the photosphere will be located at the base of the HIF at all phases of pulsation as is the case with RRab stars \citep{kan95,kan96}. Figure \ref{fig9} also indicates that as the period gets shorter the "distance" in the mass distribution between the HIF and the photosphere is the same at maximum and minimum light. This does not necessarily mean that there will be a flat PC relation at both these phases as explained by \citet{kan95} and \citet{kan96} for the case of RRab stars. These short period, fundamental mode Cepheids will be investigated in a future paper.



\section{Light curve structure}

\begin{figure*}
       \vspace{0cm}
       \hbox{\hspace{0.2cm}\epsfxsize=8.5cm \epsfbox{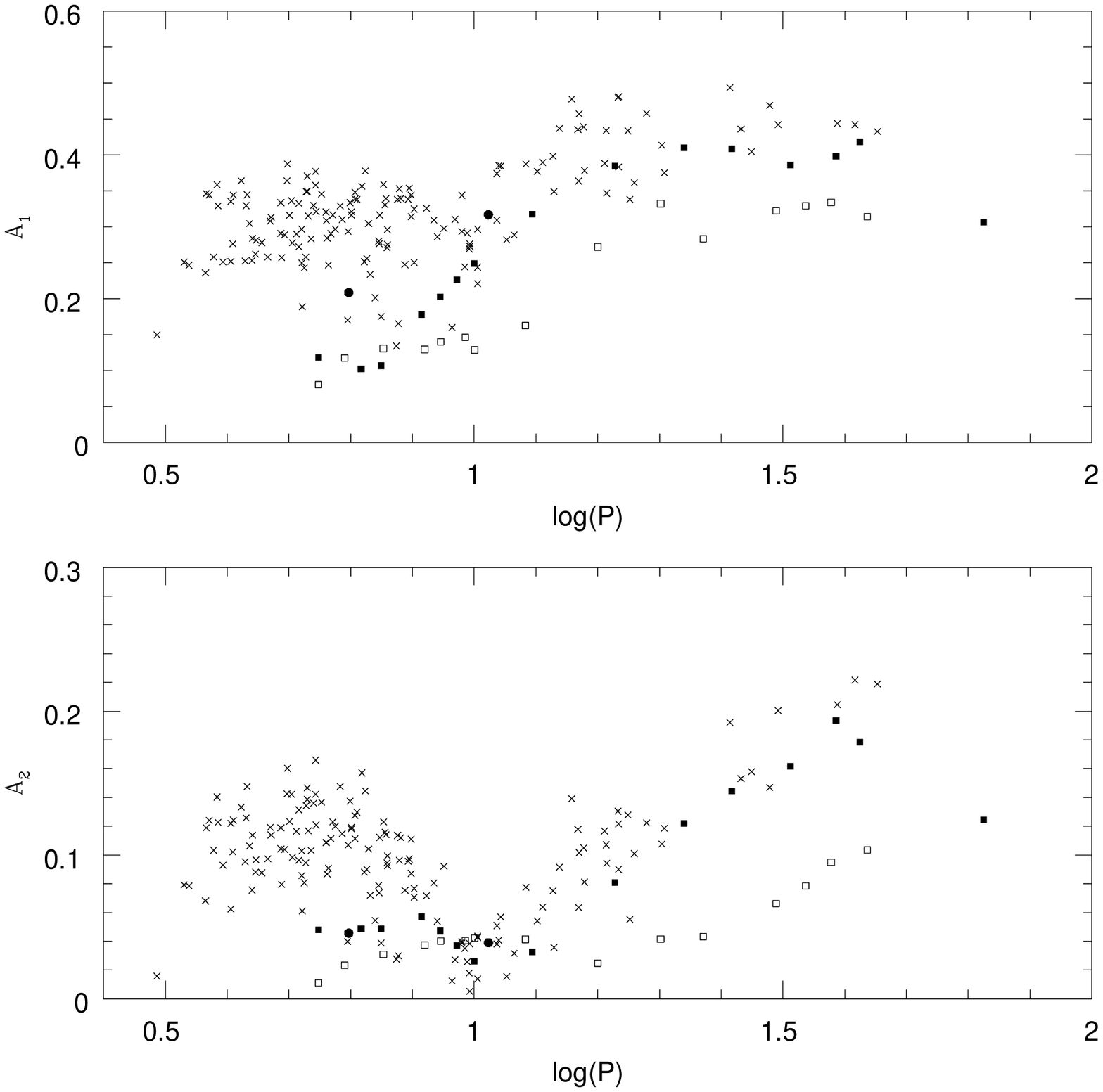}
         \epsfxsize=8.5cm \epsfbox{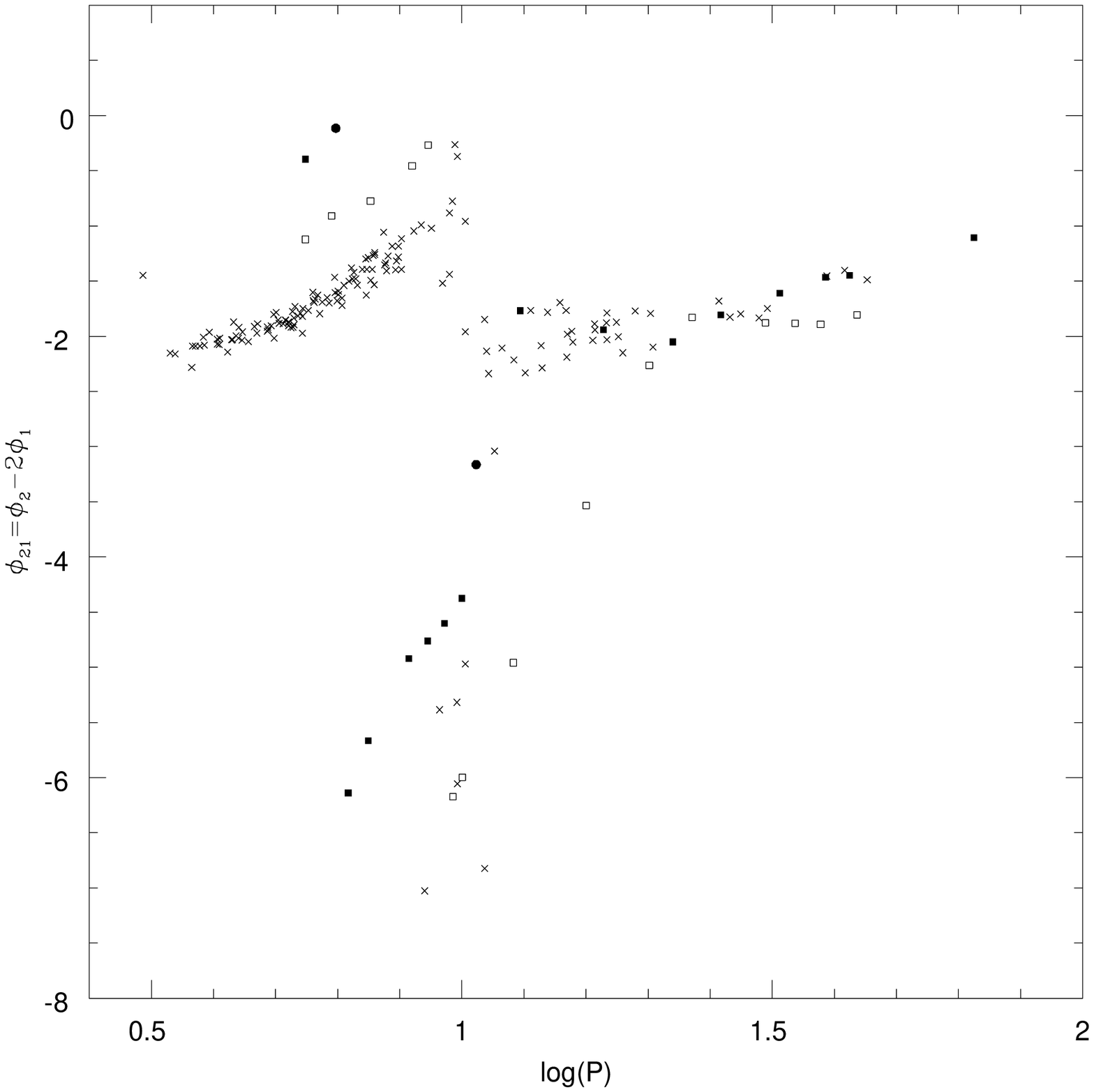}}
       \vspace{0cm}
       \caption{The plots of first two Fourier amplitudes, as defined in equation (4), and $\phi_{21}$ as function of period for the observational data and models. The symbols are same as in Figure \ref{fig1}. {\it Upper Left} (a): Plot of $A_1$ vs $\log(P)$; {\it Lower Left} (b): Plot of $A_2$ vs $\log(P)$. {\it Right} (c): Plot of $\phi_{21}$ vs $\log(P)$.}
       \label{fig10}
     \end{figure*}

     It is desirable to compare the theoretical light curves from models {\it quantitatively} to the observed light curves, for example by Fourier decomposition \citep{pay47,sim95} or Principal Component Analysis \citep{kan02}. The three panels of Figure \ref{fig10} present the plots of the Fourier amplitudes $A_1$ \& $A_2$ (top and bottom left-side panels respectively) and ${\phi}_{21} = {\phi}_2-2{\phi}_1$ (right-side panel) against $\log(P)$ for the observed data and the model light curves. The Galactic Cepheid data, which are the same as those used in KN, and the V band model light curves are subjected to a Fourier decomposition of the form:

     \begin{eqnarray}
       V = A_0 + \sum_{k=1}^{k=N}[A_kcos(k\omega t + {\phi}_k)],
     \end{eqnarray}

\ni  where $\omega = 2\pi/P$, and $P$ is the period. We used a simulated annealing technique to perform this Fourier decomposition. This method significantly reduces numerical noise in the Fourier decomposition of sparsely sampled periodic data \citep{nge03}.  

     In terms of these diagrams, there is much better agreement between models and observations when we use the Chiosi ML relation (equation [2]). These diagrams also illustrate that the agreement between models and data in the $\log (P)$-temperature or $\log (P)$-colour plane is a necessary, but not sufficient, condition for agreement between observed and theoretical light curves. This also emphasizes that the physical effect of interest in this paper - the interaction of the photosphere and HIF - is a fundamental physical effect which is, to some extent, independent of the ML relation used.

     The Florida code used in this study  contains a recipe to calculate the time dependent turbulent convection whilst SKM found a similar result using a purely radiative, dynamically zoned code. Our results imply that this interaction is also independent of numerical techniques and physics included in the code. What is needed is the presence of hydrogen and an accurate description of the opacity when hydrogen starts to ionize. 

     At first sight, the left panels of Figure \ref{fig10} might lead one to question the ML relation in equation (3). However, the numerical recipe to model convection contains seven dimensionless parameters \citep[the $\alpha$ parameters, ][]{yec99}, many of which can influence the amplitude. A more detailed study and comparison of the effect of ML relations may remove this discrepancy. For example, the two models constructed with the ML relation used in SKM, shown as black dots in Figure \ref{fig10}, do lie closer to the observations in the Fourier plane. This may also explain the discrepancy between observations and theory in the $\log (P)$-$(V-I)_{max}$ plane for $\log (P) > 1.3$ when the \citet{bon00} ML relation is used.



\section{Conclusion and discussion}

     By looking at the way the Cepheid photosphere, the region where the Cepheid continuum is generated, interacts with the HIF, we have provided a simple qualitative physical explanation for the observed PC properties of fundamental mode Galactic Cepheids with $\log (P) > 0.8$. This explanation relies on the fact that the opacity in a Cepheid becomes very high when hydrogen starts to ionize. This acts as a wall and prevents the photosphere going any deeper, and leads to a flat relation between period and temperature at the phase when the HIF interacts with the photosphere. For Galactic Cepheids this is observed at maximum light. At other phases, the photosphere is located away from the opacity wall and its temperature is related to the global properties of the star and hence its period. This explains, convincingly, why the Galactic Cepheid period-temperature relation is flat at maximum light and has a non-zero, single slope at mean and minimum light. 

     The qualitative nature of this idea remains true whether the pulsation code is purely radiative (as in SKM), or has a numerical recipe to model time dependent turbulent convection as in this work. The interaction between the photosphere and HIF may also provide some explanation for the suggestion made by \citet{ker04}, that the region where spectral lines are formed do not necessary move homologously with the region where the K-band continuum is formed. In addition, because we have used two ML relations with a wide range of $L/M$ ratio in our study, and because the same physical effect is present in models constructed with either relation, the qualitative nature of our result is independent of the ML relation used. However, the ML relation used and its variation with metallicity, will dictate how the interaction of the HIF with the photosphere changes with Cepheids in a different metallicity environment. In the next paper in this series, we will investigate how these ideas can be used to explain the PC, AC and PL properties of fundamental mode LMC Cepheids.

     We have also found evidence that the non-linear nature of the Galactic PC relation at maximum light reported by KN is due to short period fundamental mode Cepheids with $\log (P) < 0.8$. These short period stars follow distinctly different PC relations and deserve further detailed study.

\subsection{The effect on the PL relation}

     What bearing do the results of this paper have on the Cepheid PC and PL relations at mean light? It is clear that PC relations at different phases contribute to the PC relation at mean light. If we choose $(V-I)_{i}$ as the extinction corrected colour at phase $i$, then a PC relation at this phase is:

     \[ (V-I)_{i} = a_{i} + b_{i}\log (P), \]

\ni and taking the average over a pulsation period, $i=1...N$, this becomes:

     \begin{eqnarray}
     <V> - <I> = {1\over{N}}\sum_{i=1}^{i=N}a_{i} + {1\over{N}}\sum_{i=1}^{i=N}b_{i} \log (P).
     \end{eqnarray}

\ni It is clear that the average intercept and slope will be affected by their values at individual phase points such as maximum or minimum light. \citet{nge04b} compute PC relations for Galactic and Magellanic Cloud Cepheids at all phases between 0 and 1 and show that, for example for Galactic Cepheids, as the phase approaches maximum light, the slope of the PC relation becomes flatter. Since the mean light PC relation affects the mean light PL relation \citep{mad91}, our aim of understanding the reasons behind changes in the PC and AC relations at different phases has a direct bearing on understanding at least one cause of the possible variation of the mean light Cepheid PL relation from galaxy to galaxy. Below we outline how our results are pertinent to studies of the variability of the Cepheid PL relation from galaxy to galaxy, though we emphasize that much of this discussion is dependent on theoretical model and data analysis currently being undertaken. Hence the following discussion is a road-map of some aspects of our future work. 

     The evidence that shows the slope of the LMC PL relation at mean light is significantly different to the slope of the mean light PL relation in the Galaxy has been provided in \citet{tam03}, \citet{nge04} and \citet{sto04}. Furthermore, the mean light LMC PL relation suffers a change at a period of 10 days whilst the mean light Galactic PL relation data is consistent with a single slope with current data \citep{tam03,fou03,kan04,nge04,san04,sto04}.
     
     To understand the effect of the PC relation on the PL relation, consider the period-luminosity-color (PLC) relation valid at any phase:

     \[ M_v = a + b\log (P) + c(V-I). \]

\ni If the PC relation is broken, say at a period $P_0$, then we have:

     \[ (V-I) = x + y\log (P), P<P_0. \]
     \[ (V-I)=x'+y'\log (P), P>P_0. \]

\ni  Substituting these two equations into the PLC relation leads to
two PL relations:

     \[ M_v = a+cx + (b+cy)\log (P),\ P<P_0. \]
     \[ M_v = a+cx' + (b+cy')\log (P),\ P>P_0. \]

\ni  Hence changes in the slope and intercept of the PC relation have a direct bearing on the slope and intercept of the PL relation. However, the results from this paper and also from KN and \citet{nge04b} imply that the mean light PC relation is affected by changes in the PC relation at different phases during a pulsation cycle, as given in equation (5). Thus the study of PC relations at various phases impacts on the variability of the mean light PL relation 
from galaxy to galaxy. 

     In this paper, we have updated and extended the work of SKM to provide an account of a simple physical mechanism, the interaction of the photosphere and HIF, which can change the properties of PC relations for Galactic Cepheids and so affect the PL relation. Specifically, a new result of our work is that for Galactic Cepheids with $\log (P) > 0.8$, the maximum light PC relation is flat, i.e., the HIF and photosphere are engaged at maximum light. We also study the changes in AC relations because they are linked to PC relations through equation (1). A flat PC relation at maximum light leads to an AC relation at minimum light of slope $\approx -0.1$ from equation (1), as seen in Figure \ref{fig3} and Table \ref{tab5}.

     However, a Cepheid in the LMC will have a different ML relation, usually in the sense that LMC ML relations have higher $L/M$ ratios. Then in order to compare Galactic and LMC Cepheids of the same period, the LMC Cepheid needs to be hotter (KN). \citet{kan95} and \citet{kan96} found that such changes in the $(M,L,T_e)$ triplet describing the model changes the relative location of the HIF and hence the phase and the range of periods at which they interact. Our hypothesis is that for LMC models, it is only after $\log (P) > 1.0$ that the HIF and photosphere are engaged at phases around maximum light, leading to a flat PC relation. This flat PC relation for long period ($\log (P) > 1.0$) LMC Cepheids is one cause for the non-linear nature of the mean light LMC PL relation (KN). Empirically, \citet{nge04b} provide preliminary evidence that the LMC PC relation is flat only for $\log (P) > 1.0$ whereas this crossover period is $\log (P) \approx 0.8$ in the case of the Galaxy. In case of SMC Cepheids, KN also provide evidence that the maximum light PC relation is not flat even for long period Cepheids. Our contention is, using some of the results of this paper, that amplitudes in SMC Cepheids are not high enough to force an interaction between the photosphere and HIF, as the SMC Cepheids have lower amplitudes than the LMC \citep{pac00}. This again must wait confirmation from a new set of SMC models.

     The final effect on PL relations at mean light will depend on the behavior at other phases, such as minimum light. For example, preliminary calculations to be presented in a future paper imply that certain changes seen in the PC relation at minimum light (KN) may correspond to the first overtone mode becoming stable. Not surprisingly, a thorough quantitative study of this must await the analysis of LMC and SMC Cepheid models, which are currently under construction. The details of the influence of flat/non-flat PC relations on the mean PC and PL relations, the crossover period of $P_0$ that could vary from galaxy to galaxy, and the comparison of PL relations at maximum and minimum light for different galaxies are beyond the scope of this paper, and will be presented in the future papers.

     \citet{gro04} recently reported a metallicity dependence in the zero point of Cepheid PL relation at mean light. They used 34 Galactic Cepheids with individual metallicity measurements and then supplemented this sample with primarily long period Magellanic Cloud Cepheids to show the existence of a quadratic term in $\log (P)$ in the PL relation. When they used primarily Galactic Cepheids in their sample, they found no evidence of a quadratic term. They interpreted these results as being due to a metallicity dependent zero point in the PL relation. These results are also consistent with the Cepheid LMC PL relation having different slopes for long and short period Cepheids as suggested by KN and \citet{san04}. \citet{san04} also plotted amplitude-mean colour relations in luminosity bins. However, KN's AC relations were along the instability strip. It will be interesting to apply the precepts behind equation (1) and multi-phase AC relations in luminosity and/or period bins, that is, across the instability strip.

\section*{acknowledgments}

This work has been supported in part by NSF (AST 0307281) to JRB. The authors would like to thank the anonymous referee for useful suggestions. 



\begin{thebibliography}{}
\bibitem[\protect\citeauthoryear{Beaulieu et al.}{2001}]{bea01} Beaulieu, J. P., Buchler, J. R. \& Koll\'{a}th, Z., 2001, A\&A, 373, 164
\bibitem[\protect\citeauthoryear{Bono et al.}{1999}]{bon99} Bono, G., Caputo, F., Cassisi, S., et al., 1999, ApJ, 512, 711
\bibitem[\protect\citeauthoryear{Bono et al.}{2000}]{bon00} Bono, G., Caputo, F., Cassisi, S., et al., 2000, ApJ, 543, 955
\bibitem[\protect\citeauthoryear{Buchler et al.}{2004}]{buc04} Buchler, J. R., Koll\'{a}th, Z. \& Beaulieu, J. P., 2004, A\&A in press  (astro-ph/0404398)
\bibitem[\protect\citeauthoryear{Caputo et al.}{2002}]{cap02} Caputo, F., Marconi, M. \& Musella, I., 2002, ApJ, 566, 833 
\bibitem[\protect\citeauthoryear{Chiosi}{1989}]{chi89} Chiosi, C., in {\it The Use of Pulsating Stars in Fundamental Problems of Astronomy}, Ed. E. G. Schmidt, Cambridge University Press, pg. 19 
\bibitem[\protect\citeauthoryear{Code}{1947}]{cod47} Code, A. D., 1947, ApJ, 106, 309
\bibitem[\protect\citeauthoryear{Cordier et al.}{2003}]{cor03} Cordier, D., Goupil, M. J. \& Lebreton, Y., 2003, A\&A, 409, 491 
\bibitem[\protect\citeauthoryear{Feuchtinger et al.}{2000}]{feu00} Feuchtinger, M., Buchler, J. R. \& Koll\'{a}th, Z., 2000, ApJ, 544, 1056
\bibitem[\protect\citeauthoryear{Fouqu\'{e} et al.}{2003}]{fou03} Fouqu\'{e}, P, Storm, J. \& Gieren, W., 2003, in {\it Stellar Candles for the Extragalactic Distance Scale}, Ed. D. Allion \& W. Gieren, Lecture Notes in Physics Vol. 635, Springer, pg. 21
\bibitem[\protect\citeauthoryear{Freedman et al.}{2001}]{fre01} Freedman, W., Madore, B., Gibson, B. et al., 2001, ApJ, 553, 47
\bibitem[\protect\citeauthoryear{Gonzalez \& Wallerstein}{1996}]{gon96} Gonzalez, G. \& Wallerstein, G., 1996, MNRAS, 280, 515
\bibitem[\protect\citeauthoryear{Groenewegen et al.}{2004}]{gro04} Groenewegen, M. A. T., Romaniello, M., Primas, F. \& Mottini, M., 2004, A\&A, 420, 655
\bibitem[\protect\citeauthoryear{Kanbur}{1995}]{kan95} Kanbur, S., 1995, A\&A, 297, L91
\bibitem[\protect\citeauthoryear{Kanbur \& Phillips}{1996}]{kan96} Kanbur, S. \& Phillips, P. M., 1996, A\&A, 314, 514
\bibitem[\protect\citeauthoryear{Kanbur et al.}{2002}]{kan02} Kanbur, S., Iono, D., Tanvir, N. R. \& Hendry, M., 2002, MNRAS, 329, 126
\bibitem[\protect\citeauthoryear{Kanbur \& Ngeow}{2004}]{kan04} Kanbur, S. \& Ngeow, C., 2004, MNRAS, 350, 962 (KN)
\bibitem[\protect\citeauthoryear{Keller \& Mutschlecner}{1970}]{kel70} Keller, C. F. \& Mutschlecner, J. P., 1970, ApJ, 161, 217 
\bibitem[\protect\citeauthoryear{Kervella et al.}{2004}]{ker04} Kervella, P, Bersier, D., Mourard, D., et al., 2004, A\&A in press (astro-ph/0404179)
\bibitem[\protect\citeauthoryear{Koll\'{a}th et al.}{1998}]{kol98} Koll\'{a}th, Z., Beaulieu, J. P., Buchler, R., Yecko, P., 1998, ApJ, 502, L55 
\bibitem[\protect\citeauthoryear{Koll\'{a}th et al.}{2000}]{kol00} Koll\'{a}th, Z., Buchler, J. R. \& Feuchtinger, M., 2000, ApJ, 540, 468
\bibitem[\protect\citeauthoryear{Koll\'{a}th et al.}{2002}]{kol02} Koll\'{a}th, Z., Buchler, J. R., Szab\'{o}, R. \& Csubry, Z., 2002, ApJ, 573, 324
\bibitem[\protect\citeauthoryear{Kov\'{a}cs}{2003}]{kov03} Kov\'{a}cs, G., 2003, MNRAS, 342, L58
\bibitem[\protect\citeauthoryear{Lejeune}{2002}]{lej02} Lejeune, T., 2002, in {\it Observed HR Diagrams and Stellar Evolution}, Ed. T. Lejeune \& J. Fernandes, ASP Conf. Series Vol. 274, pg. 159
\bibitem[\protect\citeauthoryear{Madore \& Freedman}{1991}]{mad91} Madore, B., F., \& Freedman, W. L., 1991, PASP, 103, 933
\bibitem[\protect\citeauthoryear{Ngeow et al.}{2003}]{nge03} Ngeow, C., Kanbur, S., Nikolaev, S., et al., 2003, ApJ, 586, 959
\bibitem[\protect\citeauthoryear{Ngeow \& Kanbur}{2004a}]{nge04} Ngeow, C. \& Kanbur, S., 2004a, MNRAS, 349, 1130 
\bibitem[\protect\citeauthoryear{Ngeow \& Kanbur}{2004b}]{nge04b} Ngeow, C. \& Kanbur, S., 2004b, in-preparation
\bibitem[\protect\citeauthoryear{Paczy\'{n}sky \& Pindor}{2000}]{pac00} Paczy\'{n}sky, B. \& Pindor, B., 2000, ApJ, 533, L103
\bibitem[\protect\citeauthoryear{Payne-Gaposhkin}{1947}]{pay47} Payne-Gaposhkin, C. 1947, AJ, 52, 218
\bibitem[\protect\citeauthoryear{Ruoppo et al.}{2004}]{rou04} Ruoppo, A., Ripepi, V., Marconi, M., et al., 2004, A\&A in-press (astro-ph/0404384)
\bibitem[\protect\citeauthoryear{Sandage et al.}{1999}]{san99} Sandage, A., Bell, R. A. \& Tripicco, M., 1999, ApJ, 522, 250
\bibitem[\protect\citeauthoryear{Sandage et al.}{2004}]{san04} Sandage, A., Tammann, G. A. \& Reindl, B., 2004, A\&A in press (astro-ph/0402424) 
\bibitem[\protect\citeauthoryear{Simon, Kanbur \& Mihalas}{1993}]{sim93} Simon, N., Kanbur, S. \& Mihalas, D., 1993, ApJ, 414, 310 (SKM)
\bibitem[\protect\citeauthoryear{Simon \& Kanbur}{1995}]{sim95} Simon, N. \& Kanbur, S., 1995, ApJ, 451, 703 
\bibitem[\protect\citeauthoryear{Storm et al.}{2004}]{sto04} Storm, J., Carney, B. W., Gieren, W., et al., 2004, A\&A, 415, 531
\bibitem[\protect\citeauthoryear{Tammann et al.}{2002}]{tam02} Tammann, G. A., Reindl, B., Thim, F., et al., 2002, in Metcalfe, N. \& Shanks, T., eds., ASP Conf. Ser. Vol. 283, {\it A New Era in Cosmology}, Astron. Soc. Pac., San Francisco, p. 258  
\bibitem[\protect\citeauthoryear{Tammann et al.}{2003}]{tam03} Tammann, G. A., Sandage, A. \& Reindl, B., 2003, A\&A, 404, 423
\bibitem[\protect\citeauthoryear{Westera et al.}{2002}]{wes02} Westera, P., Lejeune, T., Buser, R., et al., 2002, A\&A, 381, 524
\bibitem[\protect\citeauthoryear{Yecko et al.}{1999}]{yec99} Yecko, P., Koll\'{a}th, Z. \& Buchler, J. R., 1998, A\&A, 336, 553
\end{thebibliography}
\end{document}